\documentclass[11pt]{article}
\usepackage{jcappub}
\usepackage{color}

\newcommand{\ds}{{\tt DarkSUSY}}
\newcommand{\mo}{{\tt MicroOmegas}}

\DeclareMathOperator{\br}{BR\!}
\DeclareMathOperator{\dif}{d\!}
\DeclareMathOperator{\e}{\mathrm e}
\def\bvec#1{\textrm{\boldmath $#1 $}}

\title{Synchrotron Emission from Dark Matter in Galactic Subhalos. A Look into the Smith Cloud}

\author[a]{N. Leite}\emailAdd{natacha.leite@desy.de}
\author[a]{R. Reuben}\emailAdd{robin.reuben@desy.de}
\author[a]{G. Sigl}\emailAdd{guenter.sigl@desy.de}
\author[b]{M.H.G. Tytgat}\emailAdd{mtytgat@ulb.ac.be}
\author[c]{M. Vollmann}\emailAdd{martin.vollmann@tum.de}

\affiliation[a]{II. Institute for Theoretical Physics, University of Hamburg, Luruper Chaussee 149, 22761 Hamburg, Germany}
\affiliation[b]{Service de Physique Th\'eorique, CP225, Universit\'e Libre de Bruxelles, Bld du Triomphe, 1050 Brusssels, Belgium}
\affiliation[c]{Physik Department T31. James-Franck-Stra\ss{}e 1, Technische Universit\"at M\"unchen, 85748 Garching, Germany}

\arxivnumber{1606.03515}

\begin{document}
  
\begin{flushright}  
{\small TUM-HEP-1048/16 
}  
\end{flushright}  
\maketitle
\newpage
\begin{abstract}

One of the key predictions of the "WIMP" paradigm for Dark Matter (DM) is that DM particles can annihilate into charged particles. These annihilations will proceed in e.~g. Galactic subhalos such as dwarf Galaxies or, as recently pointed out, high velocity clouds such as the ``Smith Cloud''. 
In this note, we focus on the radio emission associated with DM annihilations into electrons and positrons occurring in the Smith Cloud. The phenomenology of this emission is discussed in quite some detail. We argue that the uncertainties in the propagation can be captured by the typical diffusion-loss length parameter (Syrovatskii variable) but that the angle-integrated radio fluxes are independent of the propagation.
We conclude that if the Smith Cloud is indeed dominated by DM, radio signals from DM annihilation stand out amongst other messengers.  
Furthermore, low frequencies such as the ones observed by e.~g. the Low Frequency Array (LOFAR) and the next-generation Square Kilometre Array (SKA) are optimal for searches for DM in the Smith Cloud. 
As a practical application, we set conservative constraints on dark matter annihilation cross section using data of continuum radio emission from the Galaxy at $22$ MHz and at $1.4$ GHz. 
Stronger constraints could be reached by background subtraction, exploiting the profile and frequency dependence of the putative DM signal. We set stronger but tentative limits using the median noise in brightness temperature from the Green Bank Telescope and the LOFAR sensitivities. 

\end{abstract}

\part{Introduction} 
In recent years, the efforts to search for evidence for the non-gravitational interactions of dark matter (DM) have been intensified. 
A particularly appealing candidate for DM, the so-called Weakly Interacting Massive Particle (WIMP) \cite{Jungman:1995df} provides the basis for the DM search.
Indirect manifestations of its non-gravitational interactions, such as annihilation in Galactic halos,  are expected to provide measurable effects on the fluxes of several astronomical messengers such as gamma rays and radio waves.

While gamma rays are by far the most studied messengers in the context of DM (see e.~g. \cite{Bringmann:2012ez} and references therein),
their fellow radio signals can provide valuable complementary information \cite{Berezinsky:1994wva,Gondolo:2000pn,Bertone:2001jv,Aloisio:2004hy,Tyler:2002ux,
Borriello:2008gy,Regis:2008ij,Bertone:2008xr,Zhang:2008rs}. 
Radio signals associated with DM will continuously be emitted as a consequence of the accelerated gyromotion of electrons and positrons ($e^\pm$) injected by WIMP annihilations. 
This type of emission is known as synchrotron radiation as it was first characterized and observed in early synchrotron experiments \cite{Elder:1947zz,Elder:1948zz}.

In the astrophysical context, however, synchrotron radiation is typically diffuse and weaker than its gamma ray counterpart.
For this reason, early studies \cite{Berezinsky:1994wva,Gondolo:2000pn,Bertone:2001jv,Aloisio:2004hy} concentrated on the Galactic Center (GC). 
In this region, radio signals associated with DM are expected to be fairly localized owing to the $e^\pm$'s large energy-loss rates and magnetic fields near the GC. 
Flux predictions based on the methods introduced in Refs. \cite{Berezinsky:1994wva,Gondolo:2000pn,Bertone:2001jv} are very much sensitive to the properties of the DM profile at the GC. 
As a consequence, they predict upper limits on the DM annihilation cross section that are rather strong \cite{Bringmann:2009ca,Laha:2012fg,Asano:2012zv,Bringmann:2014lpa,Cholis:2014fja}. 

In this note, we instead focus on a different kind of target, namely Galactic DM subhalos.
These objects are promising for indirect DM searches mainly because their matter content is dominated by DM (see, e.~g., Ref. \cite{Mateo:1998wg}). 
In particular, negative searches for DM in a particular type of Galactic subhalos, namely dwarf spheroidal galaxies (dSphs), with the gamma ray Fermi telescope, provide the (up to date) strongest upper limits on the annihilation cross section for several masses and annihilation channels \cite{Ackermann:2015zua,Ahnen:2016qkx}.

In addition to the aforementioned dSphs, some Galactic H{\small I} substructures in the form of High Velocity Clouds (HVC) (see, e.~g., Ref. \cite{2013pss5.book..587W}) might be composed of mainly DM. 
This observation was recently pointed out in Refs. \cite{Nichols:2014qsa,2016ApJ...816L..18G} but originally put forward by \cite{1999ApJ...514..818B,2001ApJ...555L..95Q}.
The main argument essentially states that the high velocity Smith Cloud (SC) \cite{1963BAN....17..203S} must be embedded in a heavy DM halo in order to have survived its (near certain) passage through the Galactic disk.
The argument is supported by simulations where the DM component is varied and where a template for the DM profile and parameters is obtained.

In light of these remarks on the SC's DM content and, as claimed in Ref. \cite{2013ApJ...777...55H}, the presumably large magnetic fields we investigate the synchrotron emission associated with the annihilation of WIMPs in its halo. 

The phenomenology of the DM-induced synchrotron emission of Galactic dSphs has already been discussed in the literature \cite{Tyler:2002ux,Colafrancesco:2006he,Borriello:2008gy,Spekkens:2013ik,Natarajan:2013dsa,
Regis:2014tga,Colafrancesco:2014coa,Natarajan:2015hma,Beck:2015rna}. 
We revisit such analyses and adapt them to the case of the studied HVC. 
Our results comprise a study of the morphological properties of the synchrotron signal at the observationally relevant frequency of 1.4~GHz for several choices of the diffusion coefficient. 
We also describe the signal's spectra for several DM masses and annihilation channels. 
In doing this, we derive a formula that resembles the corresponding one in the gamma ray case.
Namely, we write the total synchrotron flux as the product of a term that only depends on the macroscopic properties of the DM ($J$-factor) and a term that depends on the microscopic synchrotron emission by each electron or positron produced by DM annihilation. 
Last but not least, we use our predictions and data at the 1.4~GHz and 22~MHz frequency to place upper limits on the annihilation cross sections of DM.
We also comment on the potential bounds that the Low Frequency Array (LOFAR) can put on these annihilation cross sections.

The article's structure is as follows. 
In part \ref{part:flux}, we describe the necessary theoretical foundations to obtain the desired radio fluxes. 
This part is divided into three sections. 
Namely, a brief discussion of the synchrotron spectrum of an electron in random magnetic fields in section \ref{sec:power}. 
In section \ref{sec:diffsol}, we analytically solve a suitable transport equation while in \ref{sec:flux} we derive a user-friendly equation for the total synchrotron flux. 
In part \ref{part:results}, we show a selection of signal predictions for the DM-induced synchrotron emission and discuss them. 
We further include both conservative and tentative constraints on the annihilation cross section for several final states that are derived by comparing our predictions with some data and telescope sensitivities. 
Finally, we compare these limits with other indirect DM detection bounds and discuss some prospects. 
The appendix \ref{sec:app} contains additional comparisons aimed at showing the impact of the uncertainty on the DM density profile and diffusion model.

\section{The Smith Cloud} 
\label{sec:Smith}

High velocity clouds derived their name from their atypical high velocities -- often not compatible with Galactic rotation -- and these clouds are detected by neutral hydrogen through measurements of the 21~cm line. 
Their origin is unclear and a subject of debate \cite{2013pss5.book..587W}.
Clouds around the Milky Way were likely conceived in the Magellanic Clouds and clouds at a considerable distance from our galaxy are compatible with models where their gas is infalling for the first time and no DM subhalo is present \cite{Chynoweth:2010tg, Thilker:2003mm}. 
Thus, the majority of HVCs is expected to have an almost negligible DM component but there should exist a number of HVCs with a heavy DM halo \cite{Westmeier:2008es}. 
The origin of the latter are DM subhalos that did not form stars but retained their gaseous component.

The idea that HVCs, and in particular the Smith Cloud, are supported by DM was originally suggested in the late 1990s \cite{1999ApJ...514..818B,2001ApJ...555L..95Q}, and the spatial distribution of a sub-population of these clouds and the expected dark matter clump distribution are consistent. 
Recently it was noted, however, that due to its enriched metallicity, the SC might instead be mainly composed of Galactic recycled material \cite{Fox2016}.
In this hypothesis the SC's DM component is not dominant.
Arguing which hypothesis is more realistic is out of the scope of this note. 
Without questioning the conclusions made in \cite{Nichols:2014qsa,2016ApJ...816L..18G} we will adopt the parameters reported there in computing our predictions for the DM-induced synchrotron emission at the SC.

The analysis of Lockmann et al \cite{2008ApJ...679L..21L} locate the SC at a distance of about 12.4~$\pm$~1.3~kpc from the Sun, 2.9~$\pm$~0.3~kpc below the Galactic plane. Projections of the cloud's orbit based on its velocity, cometary shape and other features indicate that it has undergone at least one passage through the Galactic plane $\sim$~70~Myr ago. This fact led to the claim in Ref. \cite{Nichols:2014qsa} that the cloud is embedded in a DM halo, since the existence of such DM halo elegantly explains the survival of the cloud's gas after passing through the Galactic plane. Based on its H{\small I} content, the cloud would have been disrupted during this passage by tidal stripping, but considering an additional DM mass $2\times 10^8 M_\odot$ the cloud could have remained gravitationally bound in the last crossing of the Galactic plane. 
Further simulations corroborate this hypothesis and in addition favor a spherically symmetrical cloud shape  \cite{2016ApJ...816L..18G}.

In Ref. \cite{Nichols:2014qsa}, hydrodynamic simulations are carried out where the gas density $n_\mathrm{H}$ and the DM mass are made variable. 
They considered gas densities $n_{\textrm H}=0.1-0.5$~cm$^{-3}$ and a DM halo following a Navarro-Frenk-White (NFW) profile \cite{Navarro:1995iw}
\begin{equation} \label{eq:NFW}
\rho(r)=\frac{r_s\rho_s}{r\left(1+\frac{r}{r_s}\right)^2}\ ,
\end{equation}
where $\rho(r)$ is the DM mass density of the cloud as a function of the distance $r$ to its center. 
Benchmark values for the normalization and scale radius of the DM halo are respectively $\rho_s=$0.57~GeVcm$^{-3}$ and $r_s=1.07$~kpc. 
These are obtained by requiring, on the one hand, that the cloud survives its passage through the Galactic disk.
This gives a lower limit on the total DM mass of the halo.
On the other hand, the halo can not be too heavy as it would have given rise to unobserved star formation. 
In Ref. \cite{2009ApJ...707.1642N} an analysis based on these considerations was done leading to parameter uncertainties $r_s=1.00-1.08$~kpc $\rho_s=0.23-0.76$~GeVcm$^{-3}$.

The distribution of mass in a DM subhalo is quite uncertain while structure formation simulations are unable to resolve scales below $\mathcal O$(0.1~kpc). The studies upon which most of this work is based, assume a Navarro-Frenk-White (NFW) type profile \cite{Navarro:1995iw} shown above.
In appendix \ref{sec:app1}, we however also consider an ``Einasto'' profile of the type \cite{1965TrAlm...5...87E}
\begin{equation} \label{eq:Einasto}
\rho_\text{Ein}(r)=\frac{\rho_s}4 \exp \left( -\frac{2}{\alpha}\left[\left(\frac{r}{r_s}\right)^\alpha-1\right]\right)\ ,
\end{equation}
where $\alpha=0.17$ and the parameters $r_s$ and $\rho_s$ are the same as for the NFW profile. Ref. \cite{2001ApJ...555L..95Q} also considers a slightly larger normalization (see \ref{sec:app}).
The magnetic field, with a peak of $\gtrsim 8\, \mu$G \cite{2013ApJ...777...55H} is even stronger than the usual Galactic field of a few $\mu$G  and measurements of its line of sight  component indicate $B_\| \geq 6 \, \mu$G \cite{McClureGriffiths:2010vp}.

\part{Radio fluxes from DM Annihilation}
\label{part:flux}
The synchrotron flux density produced by a generic distribution of isotropically-emitting emitters is given by 
\begin{equation} \label{eq:flux}
S(\nu)=\frac1{4\pi}\int\dif\Omega\cos\theta\int_{\rm l.o.s.}\dif l(\Omega)j_\nu(\bvec r)\simeq\int\dif V\frac{j_\nu(\bvec r)}{4\pi l^2} \ ,
\end{equation}
where $\nu$ is the observed frequency and ``l.~o.~s.'' stands for line of sight. Since we will be interested in sources with small angular size, we can safely neglect $\mathcal O(\theta^2)$ terms. 
The flux is characterized by the emission coefficient \citep{1992hea..book.....L}
\begin{equation}\label{eq:emicoef}
j_\nu(\bvec r) =\int\dif E P^\textrm{synch}_{\nu}(E;\bvec r) f_e(\bvec r)\ ,
\end{equation} 
where $P_\nu^\textrm{synch}(E;\bvec r)$ is the spectral power synchrotron-radiated by one electron sitting in a volume element $\dif V$ centered at position $\bvec r$ and $f_e(\bvec r)$ is the electron number density per unit energy. 

\section{Injection}
The function $f_e(\bvec r)$ can only be determined if we know how electrons are injected in a given DM (sub)halo. 
This injection is well-described by the source function
\begin{equation}
\label{eq:source}
Q_\text{DM}=\frac{\langle\sigma v\rangle}{2m_\textrm{DM}^2}\rho^2(\bvec r)\sum_\text{chann.}\br\frac{\dif Y_e}{\dif E}\ ,
\end{equation}
where $\langle\ldots\rangle$ denotes velocity ($v$) average, $\sigma$ is the annihilation cross section, $m_{\rm DM}$ is the WIMP-mass and $\dif Y_e/\dif E$ is the yield of electrons with energies $(E,E+\dif E)$ in an annihilation channel with branching ratio BR. For most DM models, the $v$ average is trivial and the DM mass-distribution $\rho(\bvec r)$ encompasses all the macroscopic features of the injection. If the DM is not self-conjugate then an additional factor of 1/2 should be included in eq. (\ref{eq:source}).

As is customary for indirect DM detection, we will consider benchmark annihilation channels -- i.~e. $b\bar b$, $W^+W^-$, $\tau^+\tau^-$, $\mu^-\mu^+$ and $e^-e^+$ --  and assume CP conservation granting that there are as much positrons as electrons produced per annihilation. The corresponding yields are tabulated functions of the electron energy and the WIMP-mass. These can be found in public software packages such as \ds{} \cite{Gondolo:2004sc}, \mo{} \cite{Belanger:2014vza} or the ``Poor Particle Physicist Cookbook for Dark Matter Indirect Detection'' \cite{Cirelli:2010xx}. We use the latter in this work. 

\section{Propagation}
\label{sec:diffsol}
The propagation is governed by a spherically symmetric stationary diffusion-loss equation:
\begin{equation} \label{eq:diffloss}
\frac{\partial f_e}{\partial t}= D(E)\frac1r\frac{\partial^2}{\partial r^2}[rf_e]-\frac{\partial}{\partial E}[-b(E)f_e]+Q_\textrm{DM}(r, E) \equiv 0 \ ,
\end{equation}
The functions $D$, $b$ and $Q_\textrm{DM}$ are, respectively, the diffusion, energy-loss and electron injection coefficients. Under our conventions they are all positive definite.
Eq. (\ref{eq:diffloss}) is the central equation in describing the electron propagation in the HVC. 
We further assume that the problem is stationary and spherically symmetric. 
\footnote{Notice that the Smith Cloud is moving with respect to the Local Standard of Rest (essentially the frame of mean motion of matter in the Galaxy) with a velocity $V_{\rm LSR}= {\cal O}(100$ km/s). On the characteristic diffusion time  scale of ${\cal O}(10^6$ yr), see Sec. \ref{sec:bdiff}, it moves on a distance ${\cal O}(1$ kpc), which is about the size of the diffusion halo, see Fig. \ref{pic:btemp}, so the validity of the stationary and spherical symmetry approximations may be questioned. A clear picture of the potential problems is given by considering the rest frame of the Smith Cloud, in which it is seen as being immersed  in a wind of (ordinary) Galactic matter of density $\rho_m$ and velocity $V_{\rm LSR}$  (we suppose here that a possible self-interaction of DM may be neglected). A relevant question is then whether the ram pressure of the wind, $p_R \sim \rho_m V_{\rm LSR}^2$,  may distort the shape of the diffusion halo produced by DM annihilation? Notice that if the wind velocity is constant, the problem is still stationary, only isotropy may be lost. Here we assume that we may neglect a possible wind gradient. This should be a reasonable approximation as long as the Smith Cloud is away from the Galactic arms, which is the case.  To address the isotropy issue, the relevant quantity to compare the effect of ram pressure is the pressure of the magnetic field, $p_B \sim B^2/2$, which confines the electrons and positrons in a halo within the Smith Cloud. As argued in \cite{2013ApJ...777...55H, McClureGriffiths:2010vp}, the magnetic pressure is indeed larger than the ram pressure, so spherical symmetry should be a good approximation. }

\subsection{The Syrovatskii variable} 
\label{sec:Syrovar}
Provided $D$, $b$ and $Q$ do not depend on $r$, eq. (\ref{eq:diffloss}) can be rewritten in the form of a heat equation
\begin{equation}
\label{eq:heat}
\frac{\partial}{\partial\lambda}\tilde f(r,\lambda,\tau)-\frac1r\frac{\partial^2}{\partial r^2}[r\tilde f(r,\lambda,\tau)]=\tilde Q(r,\lambda)\ ,
\end{equation}
with $\tilde f\equiv bf_e$, $\tilde Q=bQ/D$ and the Syrovatskii variables $\lambda\equiv\lambda(E)$ and $\tau(t,E)$ are defined by \cite{1959AZh....36...17S,MartinVollmann:2015gea}
\begin{equation}
\label{eq:syrova}
\dif\lambda=-\frac{D(E)\dif E}{b(E)}\quad ,\quad \dif\tau=\dif t-\frac{\dif E}{b(E)}\ .
\end{equation}

Analytical solutions to the heat equation are obtained by means of the Green-function method
\begin{equation}
 \tilde f(r,\lambda)=\int\dif\lambda'\dif\tau'\dif r'\frac{r'}{r}G(r,r',\lambda,\lambda',\tau,\tau')\tilde Q(r',\lambda')\ ,
\end{equation}
where 
\begin{equation}
G(r,r',\lambda,\lambda',\tau,\tau')=\Theta(\lambda-\lambda')\frac{\e^{-\frac{r^2+r'^2}{4(\lambda-\lambda')}}}{\sqrt{\pi(\lambda-\lambda')}}\sinh\left[\frac{rr'}{2(\lambda-\lambda')}\right]\delta(\tau-\tau')
\end{equation}
is a Green's function satisfying the 1D heat equation with boundaries at infinity since the SC is embedded in the Galactic diffusion disk. 

The simplification that results from passing from eq. (\ref{eq:diffloss}) to eq. (\ref{eq:heat}) tells us that the variable $\lambda$ is the most appropriate one for this problem. 
The Syrovatskii variable has units of a squared length. 
Physically speaking, $\sqrt{\lambda(E)}$ defines the typical length scale of diffusive transport by which the energy $E$ of an electron will drop significantly\footnote{Notice that according to our sign convention in eq. (\ref{eq:syrova}) $\lambda$ is positive-definite provided the energy dependence of $b$ is harder than that of $D$.}.

Our analysis is therefore valid provided that $\lambda$ is always smaller than the squared height of Galactic disk diffusion zone. Specifically, $\sqrt\lambda\lesssim 5~$kpc in order not to conflict with Boron over Carbon constraints (see e.~g. \cite{Maurin:2001sj}).

By expressing the energy in terms of the Syrovatskii variable in eq. (\ref{eq:emicoef}) we notice that 
\begin{equation}
\label{eq:sourcelambda}
\tilde Q(r,\lambda)=\frac{\langle\sigma v\rangle}{2m_\textrm{DM}^2}\rho^2(r)\sum_\text{chann.}\br\frac{\dif Y_e}{\dif\lambda}
\end{equation}
and the emission coefficient acquires the following alternative expressions
\begin{eqnarray}
\label{eq:master}
j_\nu(r) &=& \int\dif\lambda\frac{P^\textrm{synch}_{\nu}[E(\lambda)]}{D[E(\lambda)]}\int\dif\lambda'\int\dif r'\frac{r'}{r}G(r,r',\lambda,\lambda')\tilde Q(r',\lambda')\\
 {} &=& \int\dif E\frac{P^\textrm{synch}_{\nu}(E)}{b(E)}\int\dif\lambda'\int\dif r'\frac{r'}{r}G(r,r',\lambda(E),\lambda')\tilde Q(r',\lambda')\nonumber\ .
\end{eqnarray}

By means of eq. (\ref{eq:flux}) and (\ref{eq:master}) we can describe the spectral properties of the signal. We will, however, also be interested in its morphology. This will be characterized by means of the brightness temperature defined by
\begin{equation}
kT_B=\frac{c^2}{8\pi\nu^2}\int_{\rm l.o.s.}\dif lj_\nu\ ,
\end{equation}
where $k$ is the Boltzmann constant and $c$ the speed of light.

\subsection{Energy losses and diffusion}
\label{sec:bdiff}
Generally, high-energetic cosmic ray electrons propagating in the Galaxy are subject to the following interactions that unavoidably result in energy losses: synchrotron emission, 
inverse Compton scattering (ICS) off the background (both CMB, star-light and dust-diffused IR light), Coulomb scattering and Bremsstrahlung. 

The energy-loss rate due to synchrotron emission and ICS can be written as \cite{Blumenthal:1970gc}

\begin{equation} \label{eq:syncICS}
b_{\text{sync+ICS}}(E, B, u) = \frac{4}{3} c \sigma_T (u_B+u_r) \gamma^2 
\approx 2.7 \times 10^{-23} \left( \frac{u_B+u_r}{\rm eV \, cm^{-3}} \right) \gamma^2 \,\,\, {\rm GeV} \, {\rm s}^{-1},
\end{equation}
where $\gamma=E/m_ec^2$; $\sigma_T\equiv8\pi e^4/3m_e^2c^4$ is the Thomson cross section, $u_r$ the radiation energy density which includes the contributions from the CMB ($u_{r, \rm CMB} \approx  0.26$~eV/cm$^3$) and from the interstellar radiation field, and $u_B$ is the magnetic energy density defined as
\begin{equation}
u_B = \frac{B^2}{8\pi} \approx 2.5 \left( \frac{B}{10 \, \mu {\rm G}} \right)^2 \, {\rm eV} \, {\rm cm}^{-3}.
\end{equation}

For collisions with thermal electrons, the Coulomb term is given approximately by \cite{1979ApJ...227..364R} 
\begin{equation} \label{eq:Coul}
b_{\text{Coul}}(E,n_e) \approx 2.7 \times 10^{-16} \, \left( \frac{n_{\rm e}}{1 \, {\rm cm}^{-3}} \right) (6.85 + \ln \gamma) 
\,\,\, {\rm GeV} \, {\rm s}^{-1}
\end{equation}
and for the Bremsstrahlung losses we can use \cite{Sarazin:1999nz}
\begin{equation} \label{eq:Brem}
b_{\text{Brem}}(E,n_e) \approx 7.7 \times 10^{-20} \left( \frac{n_{\rm e}}{1 \, {\rm cm}^{-3}} \right) \gamma \left( 0.36 + \ln \gamma \right)   
\,\,\, {\rm GeV} \, {\rm s}^{-1}.
\end{equation}

Following Refs. \cite{Nichols:2014qsa,2013ApJ...777...55H} we use $n_e=0.5~\text{cm}^{-3}$ and $B=10~\mu$G as benchmark parameters. 
\begin{figure}
  \begin{center}
    \includegraphics[width=.6\linewidth,natwidth=610,natheight=642]{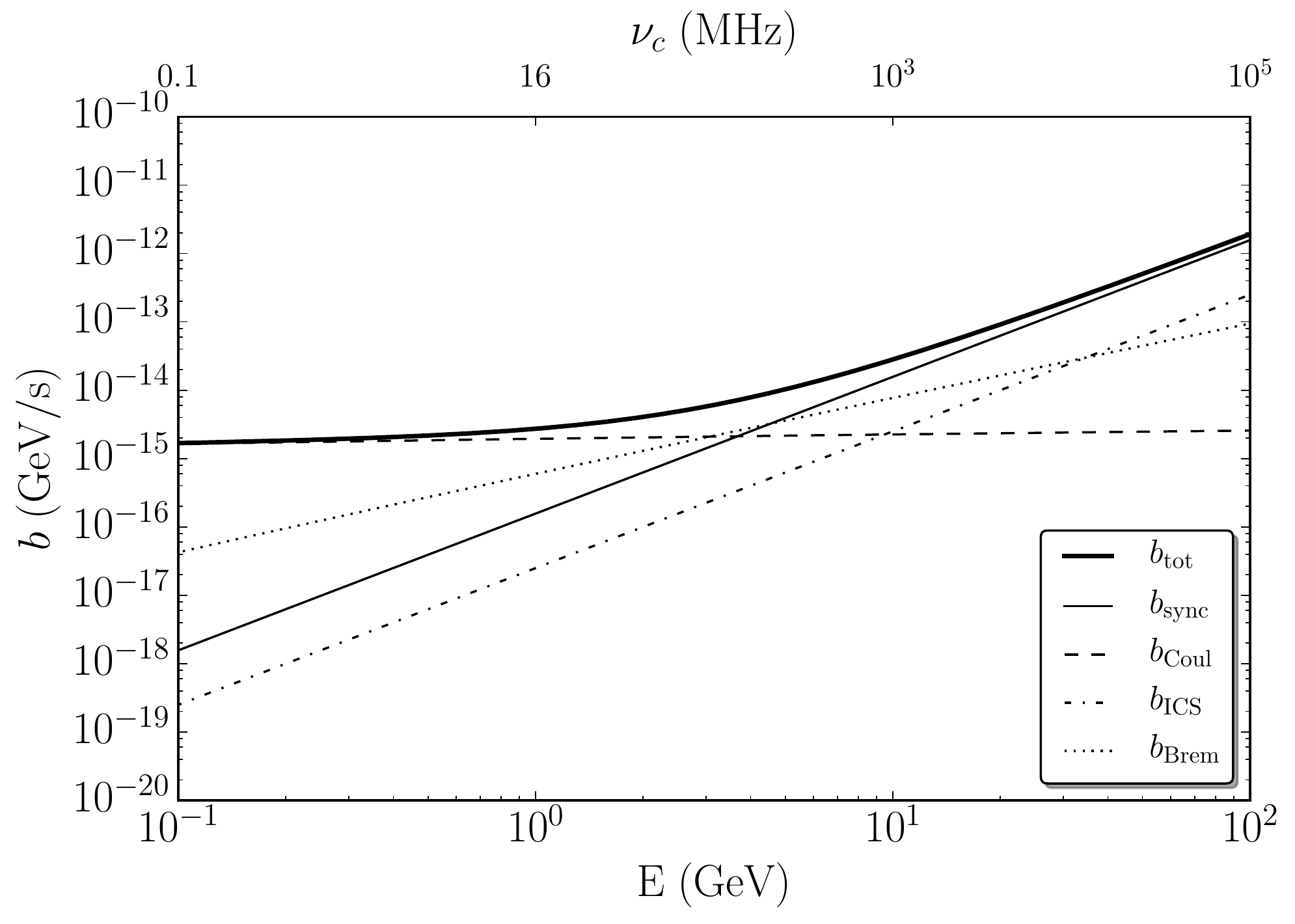}
    \caption{Energy loss rates for the different mechanisms ~\eqref{eq:syncICS},~\eqref{eq:Coul} and \eqref{eq:Brem}.} \label{pic:b}
  \end{center}
\end{figure}

In figure \ref{pic:b}, one can observe that high-energetic electrons will effectively experience energy losses associated solely with their interaction with the ambient electromagnetic field (CMB+$B$-field). Electrons with energies between $\sim$1 and 10~GeV will experience all the processes discussed equally strongly. Low energetic electron energy-loss is dominated by Coulomb interactions.

Diffusion, on the other hand, is driven by the turbulent component of the magnetic field. 
In our approximate diffusion model, we assume that diffusion follows a power law with respect to the energy, 
\begin{equation}
 D(E) = D_0\left(\frac{E}{E_0}\right)^\delta,
\end{equation}
where $D_0$ is the diffusion normalization, $E_0$ some typical energy, for instance the energy at which synchrotron losses start to dominate (we adopt $E_0=$1~GeV), and $\delta$ the spectral index.
As a consequence of this description, we observe that for high electron energies the Syrovatskii variable also follows a power law: $\lambda\propto E^{-\alpha}$ ($E\gg1~$GeV). 
In a Kolmogorov model, the turbulent spectrum has the index $\alpha=3/2$. The normalization factor
$\lambda_0$ therefore serves as a parameter that characterizes the typical length scale of diffusive transport. In other words, $\sqrt{\lambda_0}$ is the typical distance that a 1~GeV electron diffuses losing most of its energy.

\section{Synchrotron Spectrum}
\label{sec:power}
In a uniform magnetic field, the habitual synchrotron power spectrum is given by (see e.~g. \cite{1992hea..book.....L})

\begin{equation}\label{eq:psynch}
P_{\rm synch}
(E, \nu, B_\perp) = \frac{\sqrt{3}e^3}{m_e} B_\perp F(\nu/\nu_c)\ ,
\end{equation} 

with the critical frequency defined as

\begin{eqnarray} \label{eq:nuc}
\nu_c &\equiv& \frac{3}{4 \pi} \frac{e}{m_e} B_\perp \gamma^2 \\
&\simeq& 16 \left(\frac{E}{\rm GeV}\right)^2\left(\frac{B_\perp}{\mu{\rm G}}\right)\, ({\rm MHz}) \ ,
\end{eqnarray}

and

\begin{equation}\label{eq:fx}
F(x) \equiv x \int_x^{\infty} K_{5/3} (x') \dif x' \, .
\end{equation}

In Eq. \eqref{eq:psynch}, $e$ is the electron charge, $B_\perp$ is the magnetic field component perpendicular to the line of sight, $\gamma$ is the electron Lorentz factor and $K_{n}(x)$ is the modified Bessel function of order $n$.

Provided that the integral of $P_{\rm synch}$ over the frequencies results in the energy loss rate discussed in the previous section, it will be quite useful to rewrite (\ref{eq:psynch}) as 
\begin{equation}
\label{eq:synchf}
 P_{\rm synch}=\frac{b_{\rm synch}(E,B_\perp)}{\nu_c(E,B_\perp)}\tilde F(\nu/\nu_c)\left(=\frac{b_{\rm synch}(E_0,B_\perp)}{\nu_c(E_0,B_\perp)}\tilde F(\nu/\nu_c)\right)
\end{equation}
where $\tilde F(x)=\frac{9\sqrt3}{8\pi}F(x)$ in order to be normalized, i.~e. $\int\tilde F(x)\dif x=1$. In the leftmost expression of eq. (\ref{eq:synchf}), we used the fact that the prefactor of eq. (\ref{eq:psynch}) does not depend on the electron energy.
For a randomly oriented field the synchrotron power function is given by \cite{1988ApJ...334L...5G}

\begin{equation}\label{eq:psynch2}
F_{\rm rand}(x) = x^2
\left[ K_{4/3}(x)K_{1/3}(x) - \frac{3x}{5}\left(K_{4/3}^2(x)- K_{1/3}^2(x)\right)  \right]\ ,
\end{equation} 
such that $\tilde F_{\rm rand}(x)=\frac{27\sqrt3}{4\pi}F_{\rm rand}(x)$. We will adopt this spectral shape in this note.

\section{Integrated flux formula}
\label{sec:flux}
In the point source approximation, formula (\ref{eq:flux}) becomes
\begin{equation}
S(\nu)\simeq\frac1{4\pi d^2}\int\dif Vn_e(\bvec r)\epsilon_\nu(\bvec r)=\frac1{4\pi d^2}\int\dif E P^\textrm{synch}_{\nu}(E)N_e(E)\ ,
\end{equation}
where $N_e(E)=\int\dif Vf_e$ is the total number of electrons with energies between $E$ and $E+\dif E$ that are inside the cloud. 
Gauss' theorem allows us to write a simple equation for $N_e(E)$, namely
\begin{equation}
\label{eq:totalNe}
\frac{\dif}{\dif E}[b(E)N_e(E)]=-4\pi\int\dif rr^2Q_\textrm{DM}(r, E)-\underbrace{4\pi R_d^2D(E)\nabla_rf_e|_{r=R_d}}_\textrm{escape rate}\ .
\end{equation}

Notice that the surface term on the right hand side of this equation is the only term that depends on the diffusion model. For energies such that the diffusion volume is large, i.~e. $R_d\gg\lambda(E)$, we can neglect that term. By plugging in eq. (\ref{eq:source}) for $Q_{\rm DM}$, we obtain an expression for the total flux that resembles the corresponding formula for prompt emission of gamma rays from DM annihilation:
\begin{equation}
\label{eq:radiototflux}
S(\nu)\simeq J_{\rm SC}\frac{\langle\sigma v\rangle}{8\pi m_\textrm{DM}^2}\sum_\text{chann.}\br\frac{\dif Y_{\rm radio}}{\dif\nu}  ,
\end{equation}
where the $J$-factor $J_{\rm SC}$ exactly corresponds to the one encountered in gamma ray studies of the SC. We also -- in analogy to the particle-physics electron yield -- define the electron ``radio yield''
\begin{equation}
\frac{\dif Y_{\rm radio}}{\dif\nu}\equiv\int\dif E P^\textrm{synch}_{\nu}(E)\frac{2Y_e(E)}{b(E)}=\int\dif E\frac{2f_{\rm syn}(E)Y_e(E)}{\nu_c(E)}\tilde F\left(\frac{\nu}{\nu_c(E)}\right)\ ,\
\end{equation}
which gives a measure of the radiated energy in form of radio waves with frequencies $\nu$ and $\nu+\dif\nu$ by a single annihilation. In the leftmost expression, we introduced the quantities $f_{\rm syn}(E)=b_{\rm syn}(E)/b_{\rm tot}(E)$ and $Y_e(E)=\int_E^{m_\textrm{DM}}\dif Y_e/\dif E$. These are, respectively, the fraction of energy that an electron loses by emitting synchrotron radiation respect to all its losses and the total yield of electrons or positrons with energies larger than $E$.

Formula (\ref{eq:radiototflux}) allows the user to obtain quick estimates of the total flux and directly relate it to the gamma ray counterpart. 
It should be used with care, though, as it holds only when the conditions mentioned above are fulfilled.

We notice that the result does not depend on the diffusion model at all. This is actually a completely general fact: even if we had assumed a complicated diffusion tensor in eq. (\ref{eq:diffloss}), formula (\ref{eq:totalNe}) would still be valid as long as the diffusion volume is infinite. This is not surprising because of the fact that all electrons will lose their energy inside the volume regardless of the way they diffuse. The resemblance between the gamma ray and the radio flux formulas is also not surprising as both describe fluxes of electromagnetic radiation propagating in straight lines.

\part{Results and discussion}
\label{part:results}
\section{Spectrum}
\begin{figure}[ht!]
  \begin{center}
\includegraphics[width=.49\linewidth,natwidth=610,natheight=642]{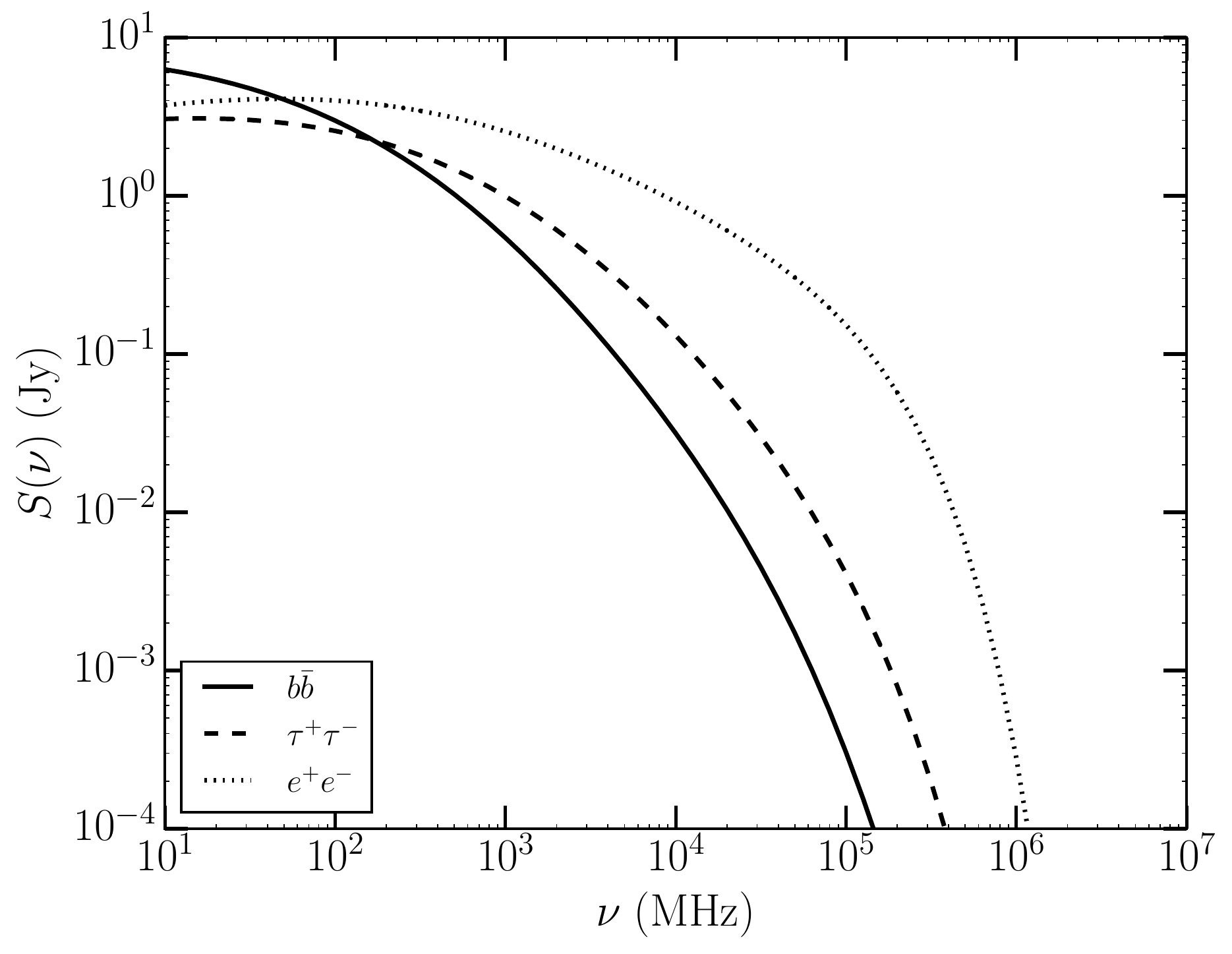} \includegraphics[width=.49\linewidth,natwidth=610,natheight=642]{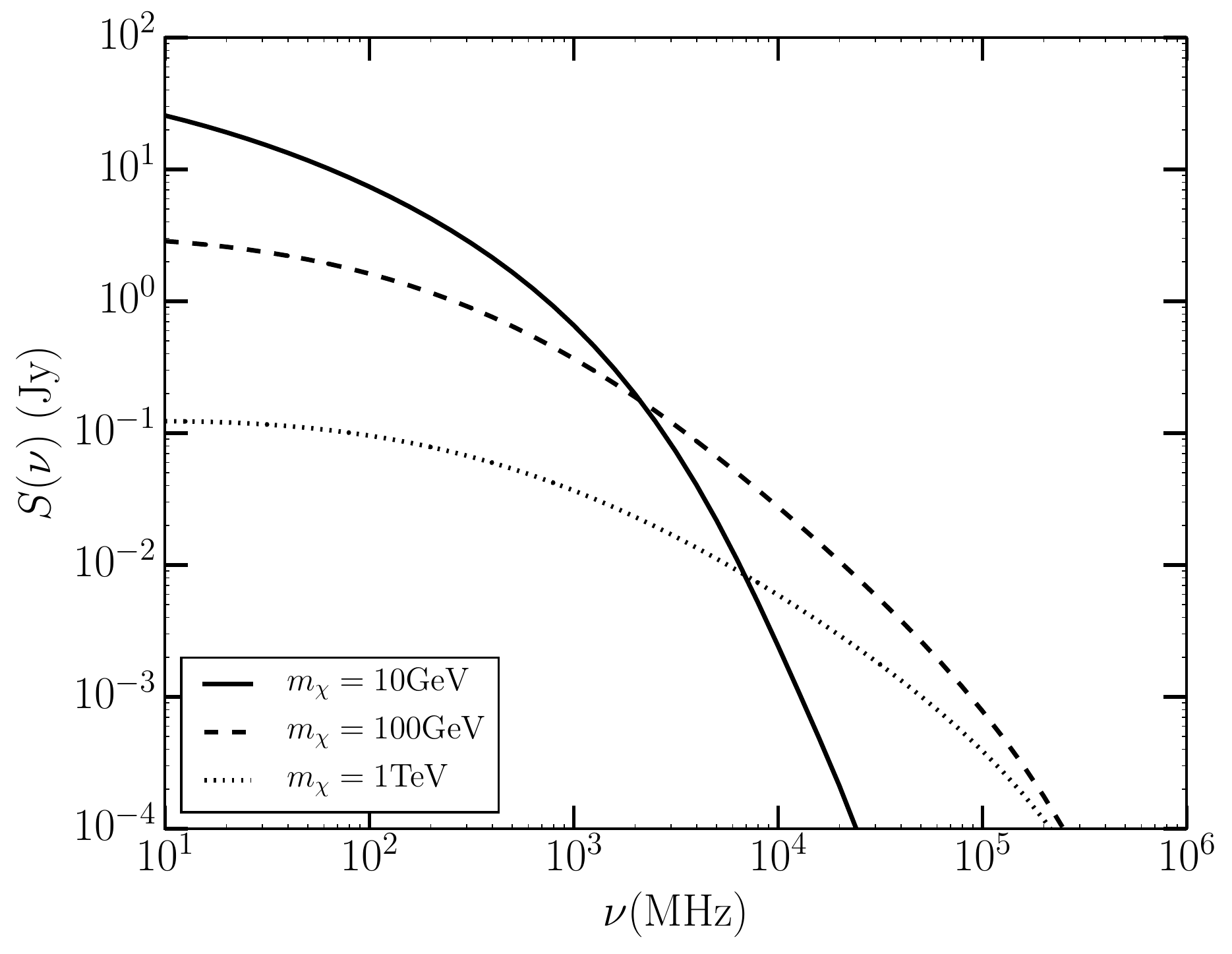}
    \caption{Synchrotron radiation flux density due to DM annihilation (with $\langle\sigma v\rangle=3\times10^{-26}~$cm$^3/$s) in the SC as a function of frequency. {\it Left}: annihilation of 50~GeV DM pairs with BR=1 into $\bar bb$, $\tau^+\tau^-$ and $e^+e^-$. {\it Right}: annihilation of 10~GeV, 100~GeV and 1~TeV DM particles into $\bar bb$ pairs.} \label{pic:totflux}
  \end{center}
\end{figure}

As argued in the previous section, provided that the volume of the SC is larger than the typical diffusion volume ($\sim\lambda^{3/2}$), formula (\ref{eq:radiototflux}) is applicable. We again assume that on average $B=10\, \mu$G \cite{2013ApJ...777...55H} with random orientations. Using this value and the hydrogen density put forward in Refs. \cite{Nichols:2014qsa,2016ApJ...816L..18G} (see fig. \ref{pic:b}) we can then obtain the radio yield for DM annihilation. We remind the reader that eq. (\ref{eq:radiototflux}) is independent of the diffusion model. 
In the two panels of fig. \ref{pic:totflux} we show radio spectra of annihilating DM with $\langle\sigma v\rangle=3\times10^{-26}~$cm$^3/$s and the $J$-factor inferred by Ref.  \cite{Nichols:2014qsa}. 

Both panels in fig. \ref{pic:totflux} confirm a feature that is general for all synchrotron signals associated with DM. 
Namely, the spectrum is quite flat at low frequencies and it has a cut-off at the characteristic frequency $\nu_c\simeq16{\rm MHz}(B/1\mu{\rm G})(m_\chi/1{\rm GeV})^2$. 
Specifically, we notice that independently of the DM mass and their leading annihilation channel, the radio spectrum is approximately constant for frequencies below $\sim$100~MHz. The signal is moreover expected to be maximal at such frequencies.
Further, the lighter the DM particle the smaller its associated characteristic frequency and the smaller the feasibility of observing it at high frequencies.

For the sake of data availability we focus in this note on the relatively large 1.4~GHz frequency that corresponds to the 21~cm wavelength. The reader should, however, keep in mind that lower frequencies are in any case more appropriate for DM radio searches.

\section{Morphology}
In contrast to the previous discussion where global properties of the electron propagation inside the SC enabled us to easily characterize the spectrum of the signal, describing the spatial properties of the signal merits more involved methods. 
These were already introduced in section \ref{sec:diffsol}. 
The main limitation in this analysis is the expected strong dependence of the results on the diffusion model. 
As mentioned there, we adopt a Kolmogorov model of diffusion $D(E)\propto E^{1/3}$ where the proportionality constant is taken as a variable.

\begin{figure}[ht!]
  \begin{center}
\includegraphics[width=.49\linewidth,natwidth=610,natheight=642]{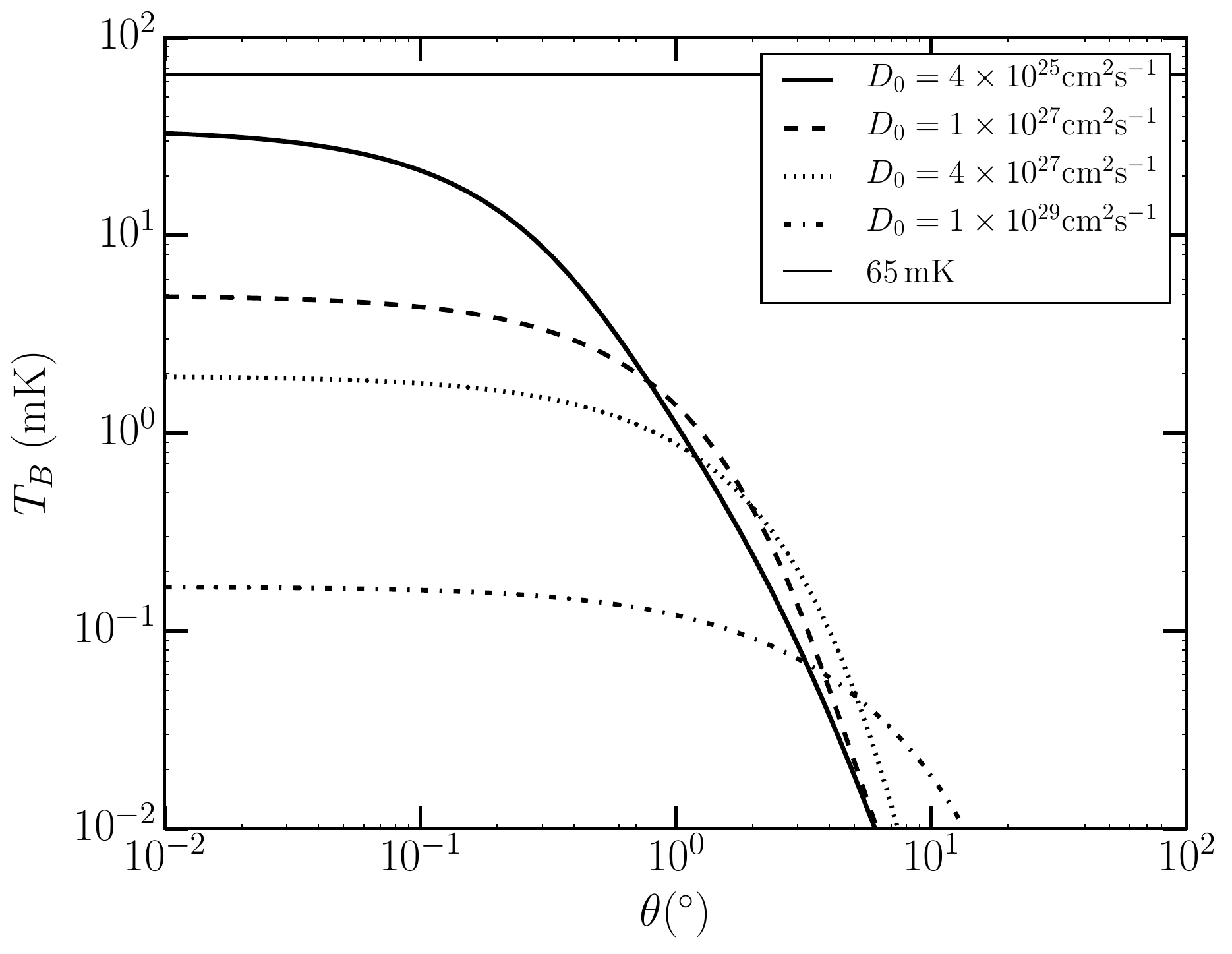} 
\caption{Signal brightness temperature at the frequency 1.4~GHz vs. observing angle for 50~GeV DM exclusively annihilating into $b\bar b$ pairs with $\langle\sigma v\rangle=3\times10^{-26}~$cm$^3/$s. Kolmogorov models for diffusion with variable normalizations $D_0$ (see sec. \ref{sec:bdiff}) were assumed.
} \label{pic:btemp}
  \end{center}
\end{figure}

In Fig. \ref{pic:btemp} the signal's brightness temperatures at $\nu=1.4$~GHz for observations centered at angles $\theta$ away from the emission center as a function of this angle are shown. For concreteness we assumed that the DM annihilation proceeds exclusively into $b\bar b$ with $\langle\sigma v\rangle=3\times10^{-26}~$cm$^3/$s and $m_{\rm DM}=50$~GeV in obtaining the curves. Each curve corresponds to a different choice of the diffusion coefficient normalization. 
The diffusion coefficients considered are chosen in such a way that the central values of the Syrovatskii variable are $\lambda_0=(0.1$~kpc$)^2$, $(0.5$~kpc$)^2$, $(1$~kpc$)^2$ and $(5$~kpc$)^2$, respectively. 

In our ``Bohr-atom model'' for the DM-induced synchrotron emission of the SC the signal form  is quite simple. Namely, the brightness temperature is essentially constant inside an angular circle whose extent depends on the normalization of the diffusion constant. 
Outside the circle the brightness temperature is exponentially cut off. 

By taking into account tidal effects of the DM subhalo and anisotropies in ambient variables e.~g. the magnetic field this picture will certainly change.
However, the property that the signal is localized and that the radius of such localization is determined by the typical diffusion length $\sqrt\lambda$ is general. 
Moreover, as a consequence of this property we notice that the radio signal is quite insensitive to the mass function of the DM subhalo (see appendix \ref{sec:app1}).

\section{Constraints on the annihilation cross section}
With the preparations made and discussed above, we are now in a position to compare our predictions with observations.
Motivated by the availability of data we focused on the 1.4~GHz frequency in the previous section. 
This frequency corresponds to the (21~cm line) radio wavelength at which the SC was first observed. 
We will also refer to some lower frequencies.
As we mentioned in the introductory sections, the SC has only been observed in surveys looking for spectral lines. 
However, the synchrotron radiation component predicted in this work has a rather broader spectrum and therefore it is relevant for radio continuum surveys.

\begin{figure}[h!]
  \begin{center}
 \includegraphics[width=.49\linewidth,natwidth=610,natheight=642]{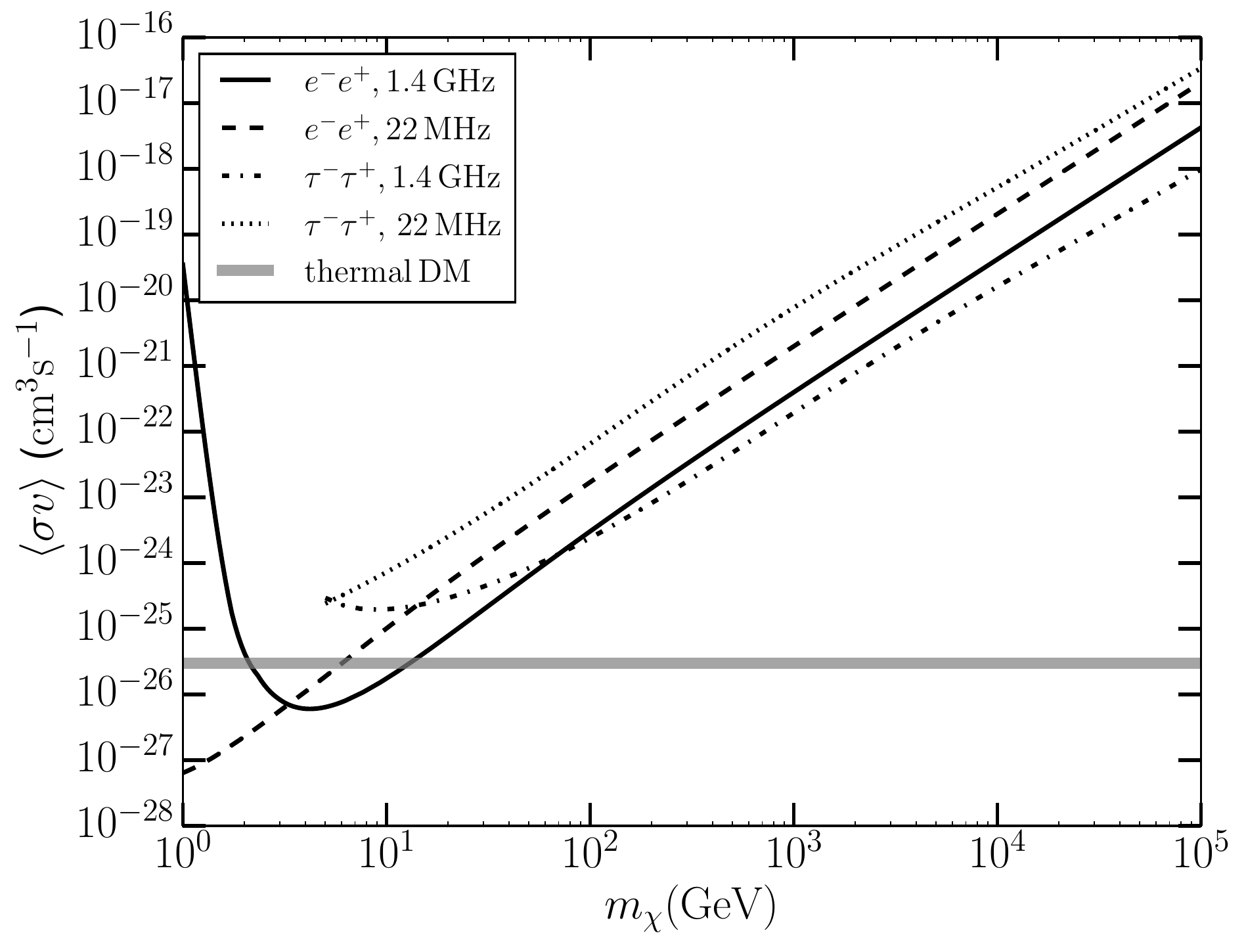} 
 \includegraphics[width=.49\linewidth,natwidth=610,natheight=642]{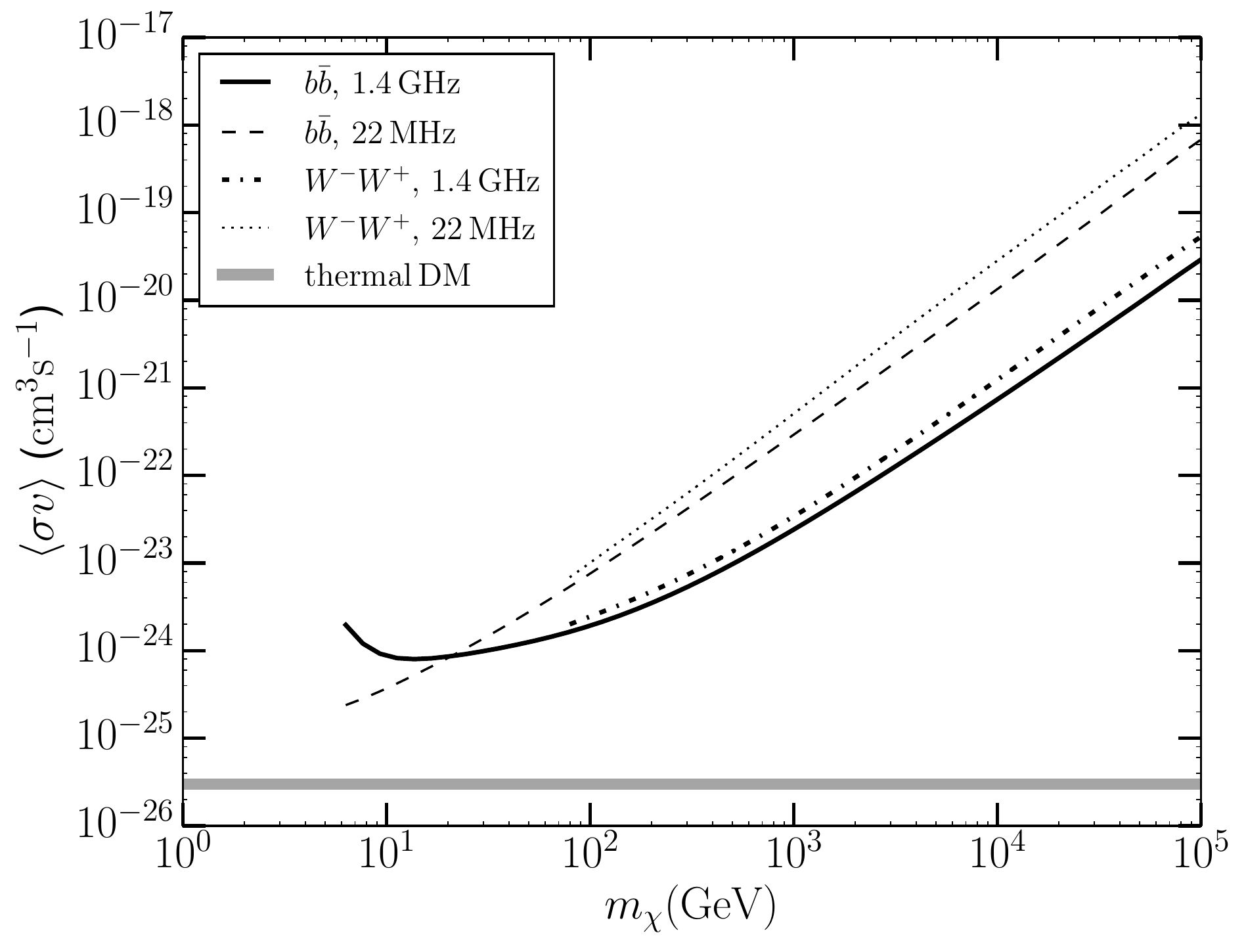}  
    \caption{Conservative limits on the DM annihilation cross section based on 22~MHz \cite{1999A&AS..137....7R} and 1.42~GHz \cite{1982A&AS...48..219R} radio continuum observations of the SC.  \textit{Left:} $e^+e^-$ and $\tau^+\tau^-$ annihilation channels; \textit{Right:} $\bar b b$ and $W^+W^-$ annihilation channels.} \label{pic:conslimits}
  \end{center}
\end{figure}

A naive approach in obtaining limits for the annihilation cross section consists of making the following reasoning. 
First, consider existing radio continuum surveys covering the SC. 
In particular, we considered data from the {\it radio continuum survey of the northern sky at 1420~MHz} \cite{1982A&AS...48..219R} as well the {\it Dominion Radio Astrophysical Observatory} (DRAO) 22~MHz \cite{1999A&AS..137....7R} which are available online on \cite{mpisurveys}.
In the position of the SC ($l\simeq39^\circ$ $b\simeq-13^\circ$) the emission is dominated by the Galactic foreground and the cloud can not be resolved. 
The non-detection of the cloud allows us to put bounds on the DM annihilation cross section.
By imposing the condition $T_\text{DM}^\text{max}<T^\text{max}_\text{fg}+2\sigma$, where $T_\text{DM}^\text{max}$ is the predicted brightness temperature evaluated at the center of the DM sub-halo, where the index ``fg'' stands for foreground and $\sigma$ for the noise level of the image, we can thus obtain rather conservative upper limits on the cross section. 
These are displayed in Fig. \ref{pic:conslimits} for $\nu=22$ MHz and 1.42~GHz.

\section{Refined searches and forecasts}
The limits shown in the previous section are based on the rather conservative assumption that the DM will be responsible for 100\% of the observed brightness at the position of the cloud. 
In other words, we did not make any assumptions on the Galactic emission, which is of course the main contributor to the total fluxes in the region where the cloud resides.  
For instance, one can make simple estimations of how the limits shown in Fig. \ref{pic:conslimits} will change if one makes more realistic assumptions as to the Galactic foreground.

Most of the Galactic emission at the aforementioned frequencies can be modelled as a simple rescaling of the Haslam 408~MHz map \cite{1982A&AS...47....1H} as a power-law $T_\text{MW}\propto\nu^{-\beta}$ with $\beta\sim2.5$ \cite{1999A&AS..137....7R}.
If one assumes that the foreground model of the SC has residuals that are $\mathcal O$(10\%)$~\times~ T_\text{fg}$ the resulting bounds on the DM annihilation cross section are one order of magnitud stronger that the ones reported in Fig. \ref{pic:conslimits}.

\subsection{Tentative limits from the GBT 1.4~GHz data}
An even more daring method in obtaining limits for the DM annihilation cross section is based on exploiting the thorough data reduction of the Green Bank Telescope (GBT) observations performed in Ref. \cite{Nichols:2014qsa,2008ApJ...679L..21L}. 
There, a median noise level of $\sim65$~mK was achieved, at the (line) frequency of maximal H{\small I} emission, though.
However, due to the relatively moderate signal-to-noise ratios encountered in the aforementioned GBT SC H{\small I} images  \cite{2008ApJ...679L..21L,Nichols:2014qsa}, we assume that the only-noise images of the SC at those slightly shifted frequencies that show no H{\small I} emission\footnote{See for instance the spectrum shown in Fig. 2 of \cite{2008ApJ...679L..21L} for $\sim-50$~km~s$^{-1}$ or $\sim150$~km~s$^{-1}$ LSR frequency shifts.}, are also characterized by the aforementioned median noise level\footnote{In our analysis we do not mean to imply that the level of accuracy of the results will be overwhelmingly good, as a more detailed analysis of the data is certainly required.
The reader should instead regard them as the best limits one can put on the annihilation cross section $\sigma v$ using the \cite{Nichols:2014qsa,2008ApJ...679L..21L} GBT 1.4~GHz data cube.}. 

\begin{figure}[h!]
  \begin{center}
 \includegraphics[width=.49\linewidth,natwidth=610,natheight=642]{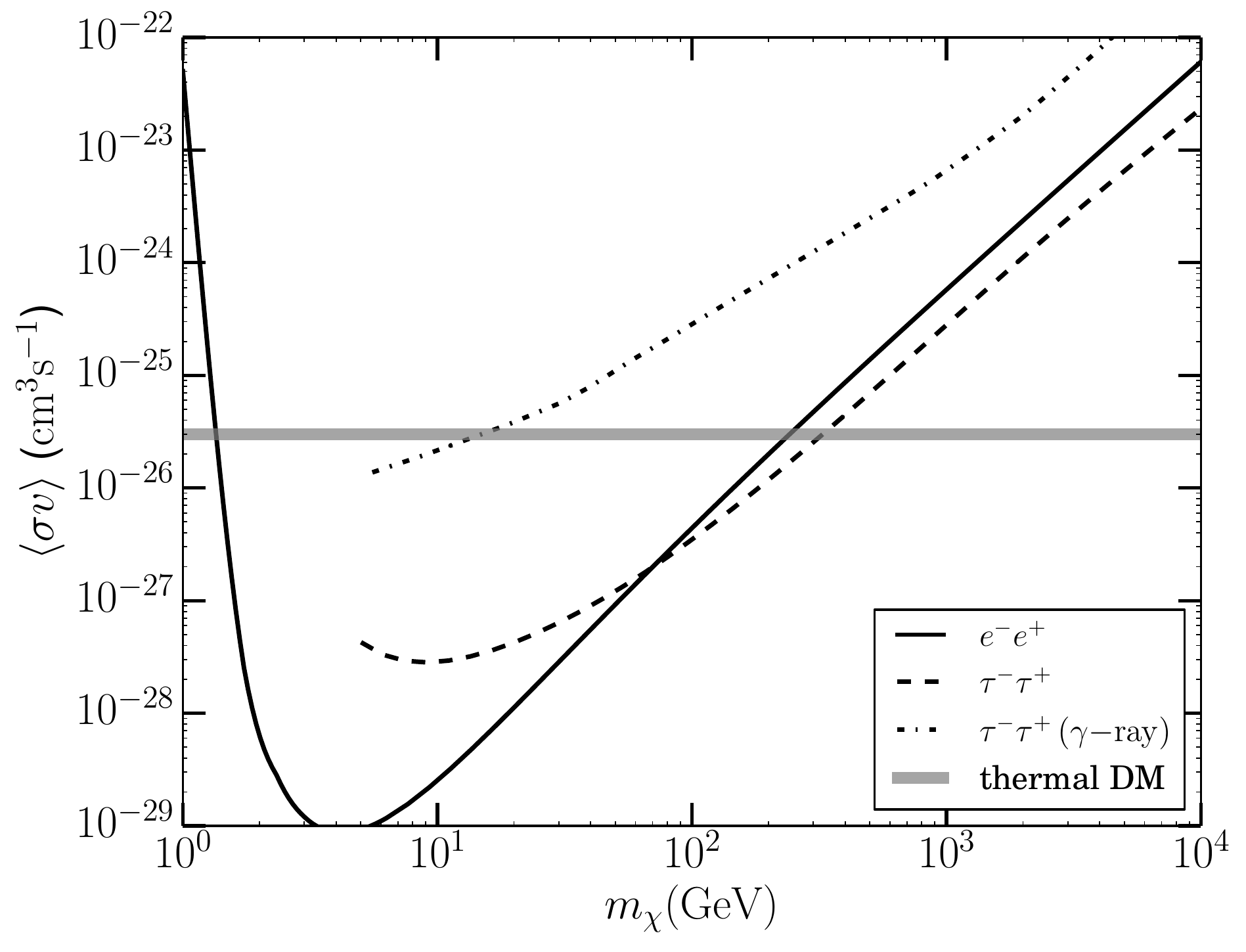} 
 \includegraphics[width=.49\linewidth,natwidth=610,natheight=642]{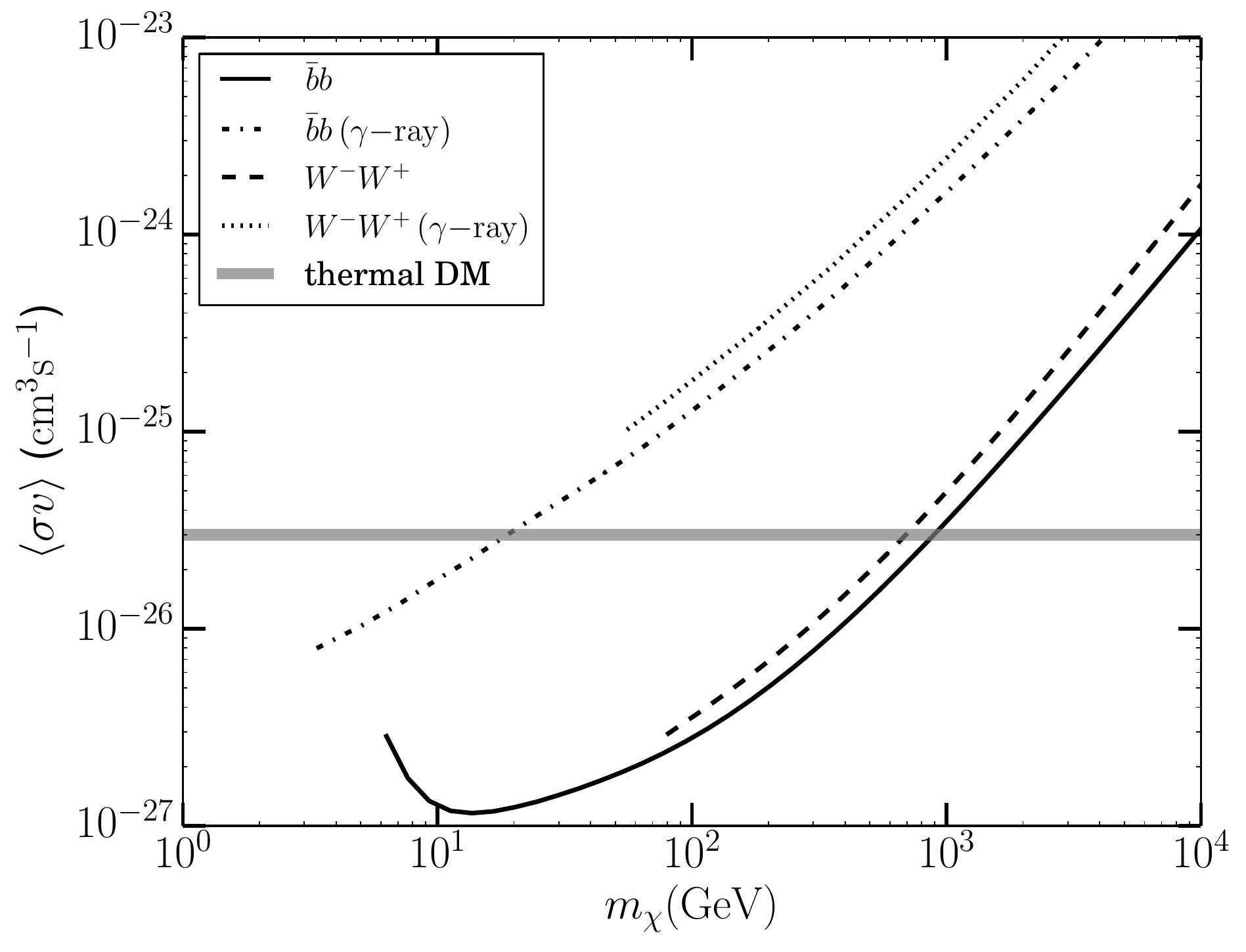}  
    \caption{Radio limits (2$\sigma$) on DM annihilation cross sections from 1.4~GHz observations of the SC. For comparison, their corresponding gamma ray limits \cite{Drlica-Wagner:2014yca} are also included. \textit{Left:} $e^+e^-$ and  $\tau^+\tau^-$ annihilation channels; \textit{Right:} $\bar b b$ and $W^+W^-$ annihilation channels.} \label{pic:limits}
  \end{center}
\end{figure}

By assuming the lack of a diffuse signal with an angular size of $\mathcal O(1^\circ)$ in the relevant data, we obtain our tentative constraints on the annihilation cross section of DM.
These are shown and compared to their gamma ray counterparts obtained in Ref.~\cite{Drlica-Wagner:2014yca} in Fig. \ref{pic:limits}. 

The criterion adopted in order to obtain tentative limits consists in comparing our predictions for $T_B(\theta=0)$ with the detectability limit of an extended signal that is sampled many times by the survey 
\begin{equation}
 T_\text{det.}\simeq\frac{65~\text{mK}}{\sqrt{N_\text{samp.}N_\text{freq.}}}\ ,
\end{equation}
where $N_\text{samp.}$ is the number of pointings that are necessary to cover the  angular extension of the DM signal and $N_\text{freq}$ is the number of independent only-noise maps in the data cube. The former can be accurately estimated by taking the ratio $\Omega_\text{sig}/\Omega_\text{beam}$ where $\Omega_\text{sig}=\pi\theta_\text{eff}^2$ is the effective DM signal solid angle from our prediction and $\Omega_\text{beam}=1.133\theta_\text{beam}^2$ if we assume that the beam is Gaussian. 
The latter, on the other hand, is determined by the amount of line-free images (variable frequency) considered in \cite{2008ApJ...679L..21L,Nichols:2014qsa}.

The effective angular size of the DM signal is given by $\theta_\text{eff}\sim\sqrt{\lambda_0}/(12$~kpc). This is confirmed in Fig. \ref{pic:btemp}, where $\theta_\text{eff}\sim1^\circ$ if $\lambda_0=(0.5~\text{kpc})^2$. It should be noted that the effective angular sizes are rather insensitive to variations of the DM mass within the considered range. The reader can convince themselves of this by looking at eq. \eqref{eq:syrova} and noticing that most of the dependence comes from the normalization of $D(E)$ while the DM mass serves just as an integration constant. 

Following what is reported in the observation \cite{2008ApJ...679L..21L}, we adopt a full width to half power (FWHP $\theta_\text{beam}$) of $\simeq3'$ and therefore $N_\text{samp.}\simeq1109\theta_\text{signal}^{\circ2}$. 
Additionally, we assume that $\sim$~2/3 out of the 485 sampled frequencies in \cite{2008ApJ...679L..21L} are interesting. 
This yields $N_\text{freq}=323$.
Specifically, we are interested in ``images'' of the SC with velocities (frequencies) that are larger either than 150~km~s$^{-1}$ or smaller than $\sim-10$~km~s$^{-1}$ LSR. As the reader can see in e.~g. Fig. 2 of Ref. \cite{2008ApJ...679L..21L} the H{\small I} line emission of the SC is subleading at those frequencies.

For completeness, in appendix \ref{sec:app2} we consider limiting cases for the diffusion normalizations in order to show that the smaller the diffusion coefficient, the stronger the constraints are -- opposite to the typical constraints expected from positrons and antiproton signals, which get stronger with the increase of the diffusion coefficient. 
We also re-evaluated the flux densities shown in Fig. \ref{pic:totflux} and the limits of Fig.  \ref{pic:limits} with $B=1~\mu$G instead of $B=10~\mu$G . 
The resulting fluxes and limits (omitted for brevity) become roughly one order of magnitude weaker.

We notice that for $m_\chi\lesssim$~10~GeV our 1.4~GHz limits are exponentially weakened. 
This is particularly apparent in the left panel of Fig. \ref{pic:limits}. 
The weakening feature is certainly a manifestation that the signal spectrum (Fig. \ref{pic:totflux}) has a cut-off at a frequency of the order of $\nu_c\simeq16{\rm MHz}(B/1\mu{\rm G})(m_\chi/1{\rm GeV})^2$.
Consequently, the lighter the DM particle, the lower the associated synchrotron cut-off frequency is.
The 1.4~GHz frequency is much too high for the radio signal of annihilating $\sim$10~GeV DM to be relevant.
Notice also that our constraints on heavy DM ($m_\chi\gtrsim$~10TeV) have a softer mass dependence than their gamma ray fellows.
This also reflects the fact that the 1.4~GHz frequency is inside the ``flat'' (optimal) regime of the signal spectrum of annihilating heavy DM.
Limits derived from high frequency maps (such as 1.4~GHz) should therefore be considered relevant for DM particles of heavier masses.

Fig. \ref{pic:limits} also demonstrates that even though the studied frequency is not the optimal one for DM searches, a thorough analysis of the GBT 1.4~GHz data can put limits on $\sigma v$ that are much better than the ones Fermi-LAT places using dSph as targets \cite{Nichols:2014qsa,Drlica-Wagner:2014yca}. 
Again, we stress the fact that we made strong assumptions on the systematics associated to the data reduction. 

\subsection{Projections for LOFAR}
Fig. \ref{pic:btemp} indicates that the sub-GHz frequency range is optimal for DM searches with radio probes. Fortunately, such frequency ranges will be probed by large-scale experiments such as LOFAR \cite{2013A&A...556A...2V} and the upcoming SKA \cite{SKA}.

Since no dedicated study of the SC in the sub-GHz range is available, we content ourselves as before by providing projected LOFAR limits on the annihilation cross section of DM. 
Specifically, we consider the (image noise) sensitivities quoted in table B.3 of Ref. \cite{2013A&A...556A...2V} and multiply them by the correction factor $1+T_\text{fg}/T_\text{sys}$. This factor accounts for the additional contribution from the Galactic foreground to the system temperature of the LOFAR antennas \cite{2012tra..book.....W,1999ASPC..180.....T}. 
We estimated $T_\text{fg}$ from existing surveys \cite{mpisurveys} as the maximum brightness temperature in a $3^\circ\times3^\circ$ square centred at the SC. In all the cases considered the correction is however rather small $\mathcal O(1\%)$.

\begin{figure}[!ht]
  \begin{center}
  \includegraphics[width=.49\linewidth,natwidth=610,natheight=642]{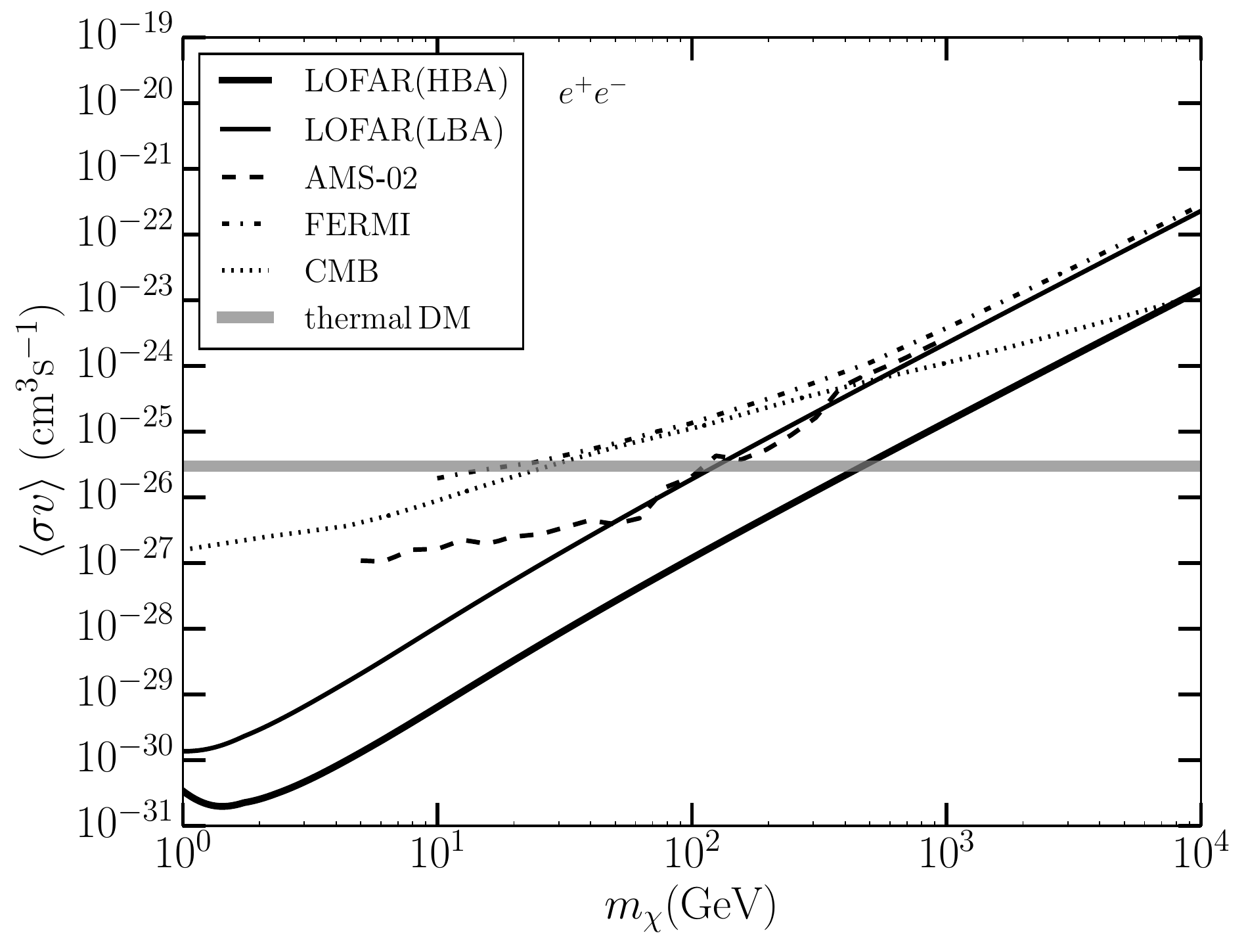}
  \includegraphics[width=.49\linewidth,natwidth=610,natheight=642]{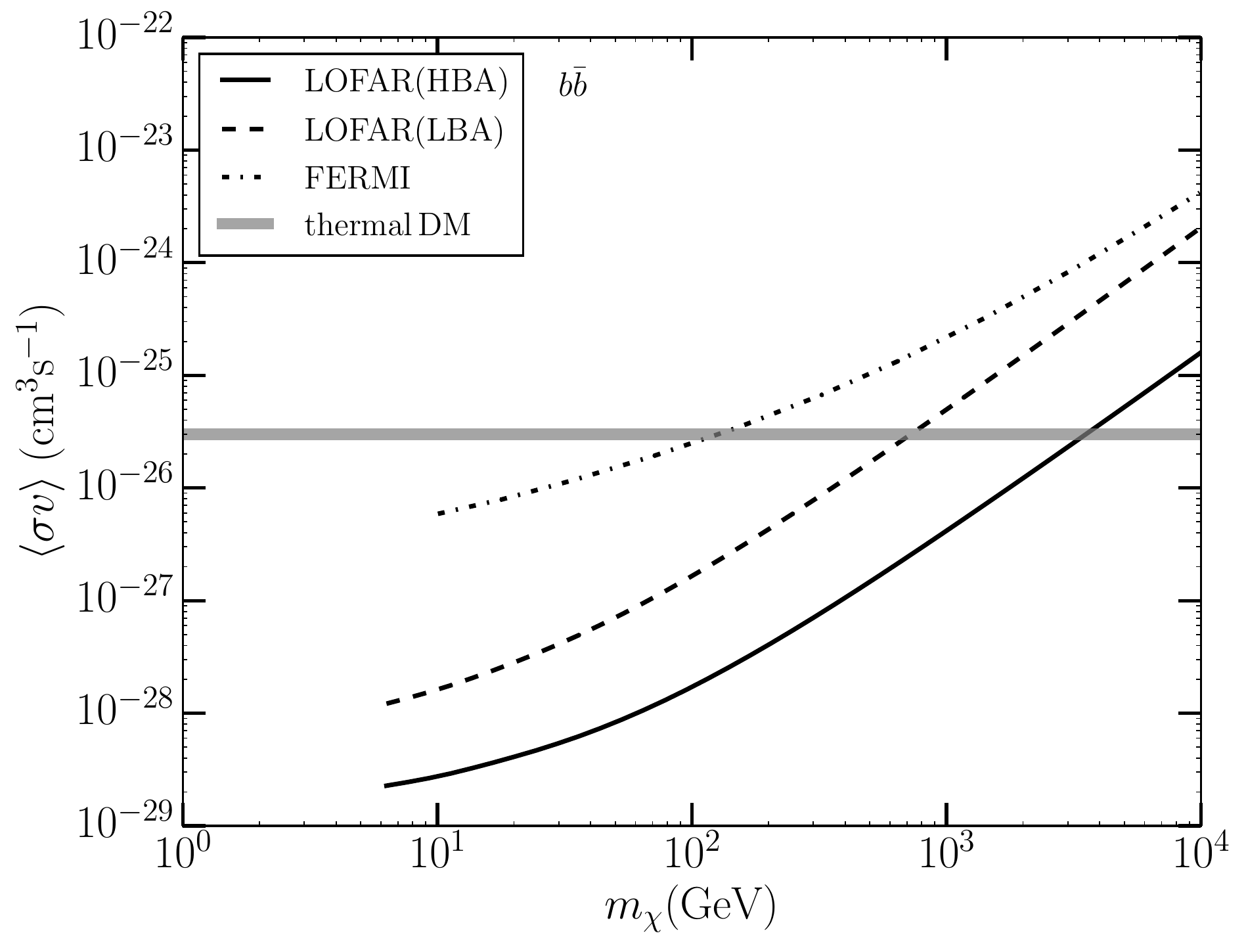}
    \caption{Projected LOFAR limits (2$\sigma$) on the DM annihilation cross section using the SC as a target. 8hrs of observation time and a typical beam size of 25$''$ are assumed at at frequency of 60 MHz (LBA) and 150 MHz (HBA). Comparison with annihilation cross section constraints drawn using data from AMS-02 \cite{Ibarra:2013zia}, dwarf spheroidal galaxies of the Milky Way observed by FERMI-LAT\cite{Ackermann:2013yva} and the CMB \cite{Kawasaki:2015peu}. \textit{Left:} $e^+e^-$ annihilation channel; \textit{Right:} $b\bar{b}$ annihilation channel.} \label{pic:lofar}
  \end{center}
\end{figure}

In Fig. \ref{pic:lofar} we show the projected LOFAR limits on the DM annihilation cross section that 60~MHz (Low Band Antenna) and 150~MHz (High Band Antenna) LOFAR measurements would provide at 95\% confidence level if the SC is not detected. These limits assume 8~hrs of integration with an effective bandwidth of 4~GHz and a typical beam size of 25$''$ in both cases. 

In a similar way as we did in the previous section we set the projected limits by comparing our predictions with the detectability limiting temperature for an extended source 
\begin{equation}
 T_\text{det.}\simeq\frac{c^2}{2k\nu^2}\frac{\Delta S_\nu}{\Omega_\text{beam}\sqrt{N_\text{samp}}}=\frac{c^2}{2k\nu^2}\frac{\Delta S_\nu}{\sqrt{\Omega_\text{beam}\Omega_\text{sig}}}\ ,
\end{equation}
where $\Delta S_\nu$ is the LOFAR sensitivity for the frequency $\nu$, which we extract from table B.3 from Ref. \cite{2013A&A...556A...2V}. $\Omega_\text{beam}$, $\Omega_\text{sig}$ and $N_\text{samp}$ were already defined in the previous section. 

Although the resulting limits are indeed stronger than the ones obtained in the previous sections, we see that the improvement is mild. 
Notice, however, that the way these limits are obtained is quite naive.
On the one hand, it assumes that the data reduction is optimal which is rather unrealistic specially for the lowest frequencies. 
In particular, we observe that the signal enhancement in going from 150~MHz to 60~MHz does not overcome the corresponding loss in sensitivity and therefore, the 150~MHz limits are better.
On the other hand, the flatness of the spectrum at sub-GHz frequencies calls for a rather multi-wavelength signal-correlation study that will lead to stronger limits. 
Also in Fig. \ref{pic:lofar} we can see the comparison between our projections and the limits set with PLANCK data throught the CMB \cite{Kawasaki:2015peu}, with the FERMI-LAT data through dShps studies \cite{Ackermann:2013yva} and with the positron flux derived from AMS-02  \cite{Ibarra:2013zia}. It should be stressed that the capability of LOFAR posing such strong constraints when looking at the SC relies on an ideal understanding of the systematics and subsequent subtraction of back- and fore- grounds.

\section{Conclusions}\label{sec:concl}
In this article we considered the solid hypothesis that the Smith Cloud is supported by dark matter. 
We argued that due to its vicinity, amount of dark matter and magnetic field strength, the Cloud is an excellent target for indirect detection of dark matter with radio data.
Furthermore, the location as well as the geometrical properties of the Cloud make the relevant phenomenology, namely the description of the synchrotron emission induced by dark matter annihilation, quite simple. 

In our Bohr-atom-like, semi-analytical model for the Smith Cloud's synchrotron signal we were able to learn about the spectral features of the signal. 
Specifically, we concluded that, for DM masses in the range 1-100~GeV, the synchrotron spectrum is quite flat in the sub-GHz regime. 
At larger frequencies the signal then decreases exponentially. 
These conclusions are quite robust as they do not depend on the diffusion model.

In contrast to the situation encountered in similar studies that consider instead the Galactic Center as their target, our results are rather independent of the DM profile. 
Instead, they mainly depend on the loss-diffusion volume defined by the Syrovatskii variable.

To obtain some first radio limits on the DM annihilation cross section we used different approaches, from conservatively admitting that the DM signal should only not overshoot the present observations to optimiscally considering that the astrophysical background is understood and/or subtracted. The former yields rather weak and the latter rather strong constraints. 
In particular we considered data from continuum radio surveys at 1.42~GHz and at 22 MHz.

An alternative approach consists of considering the noise level of a reduced image of the Smith Cloud using GBT's 21 cm observations.
In this case, constraints that are even stronger than the ones reported by the Fermi-LAT gamma-ray telescope in their searches for DM using dwarf galaxies were obtained.

The presented cross-sections serve as limiting beacons between which the realistic DM signal should fall. 
Further studies to understand the foregrounds involved and to be able to undertake a suitable subtraction of the astrophysical background and galactic foreground are therefore most necessary. 

Motivated by the fact that data in the sub-GHz band will be the best suited for radio searches, we also presented predictions for the constraints that could be set on the DM with the Low Frequency Array LOFAR. 
In particular, we considered the Low and High band antennas, at 60 MHz and 150 MHz frequencies respectively. 
The obtained projections correspond to a best-case scenario since they assume an optimal data reduction, but despite this they indeed provide constraining power comparable to other methods.

In summary our results 
favour multi-wavelength searches. 
According to them, the search for morphological correlations using a set of frequencies that are orders of magnitude apart can be quite effective.
They also intend to show that the study of HVCs is of large interest for indirect dark matter searches and especially the potentials that future radio surveys offer in this respect.

\section*{Acknowledgments}
The work of N.L., R.R. amd G.S. was supported by the ''Helmholtz Alliance for Astroparticle Physics (HAP)'' funded by the Initiative and Networking Fund of the Helmholtz Association.  The work of M.T. by the Fonds National de la Recherche Scientifique and by the Belgian Federal Science Policy Office through the Interuniversity Attraction Pole P7/37 and the work of M.V. by DFG Cluster of Excellence ``Origin and Structure of the Universe''. We would like to thank Marcus Br\"uggen, Maria Vittoria Garzelli and the anonymous referee for their very valuable hints as well as Carmelo Evoli and Torsten Bringmann for their early participation in a similar project. G.S. and M.V. would also like to thank Georg Raffelt and his group for their hospitality. An essential part of this work was completed at the Werner Heisenberg Institute in Munich. One of us (M.T.) would like to thank J\'er\^ome Vandecasteele and Antoine Pasternak for useful discussions. M.V. would also like to thank Francesco de Gasperin and Roberto Lineros for their comments on the final draft.

\part{Appendix} 
\label{sec:app}
\appendix
\section{Alternative DM distributions}\label{sec:app1}
Up to now, we displayed spectra and limits that result only from the Navarro-Frenk-White (NFW) ansatz for the SC \eqref{eq:NFW} given by Ref. \cite{Nichols:2014qsa}. However, for completeness, we also considered different choices.

\begin{figure}[!ht]
  \begin{center}
 \includegraphics[width=.49\linewidth,natwidth=610,natheight=642]{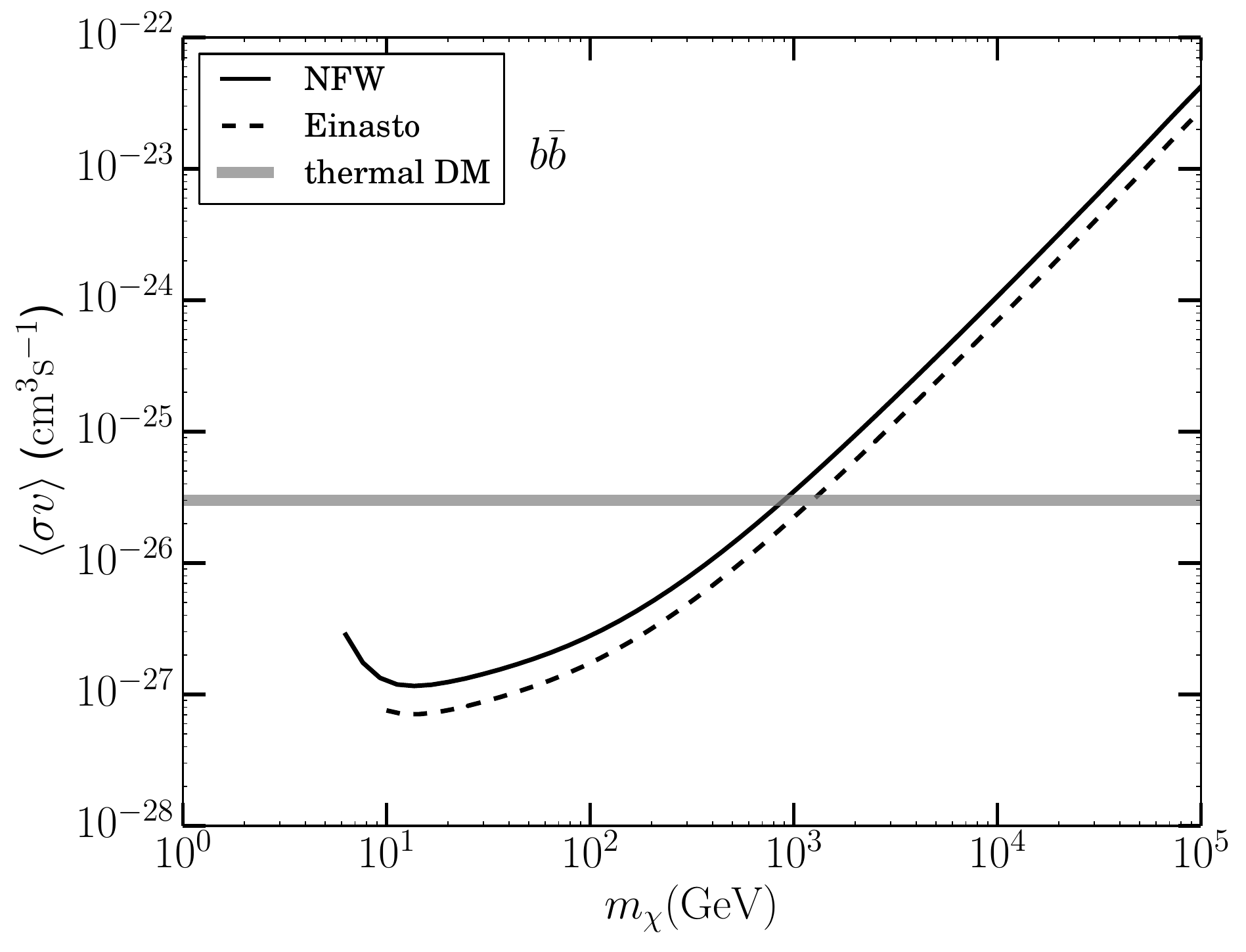}
 \includegraphics[width=.49\linewidth,natwidth=610,natheight=642]{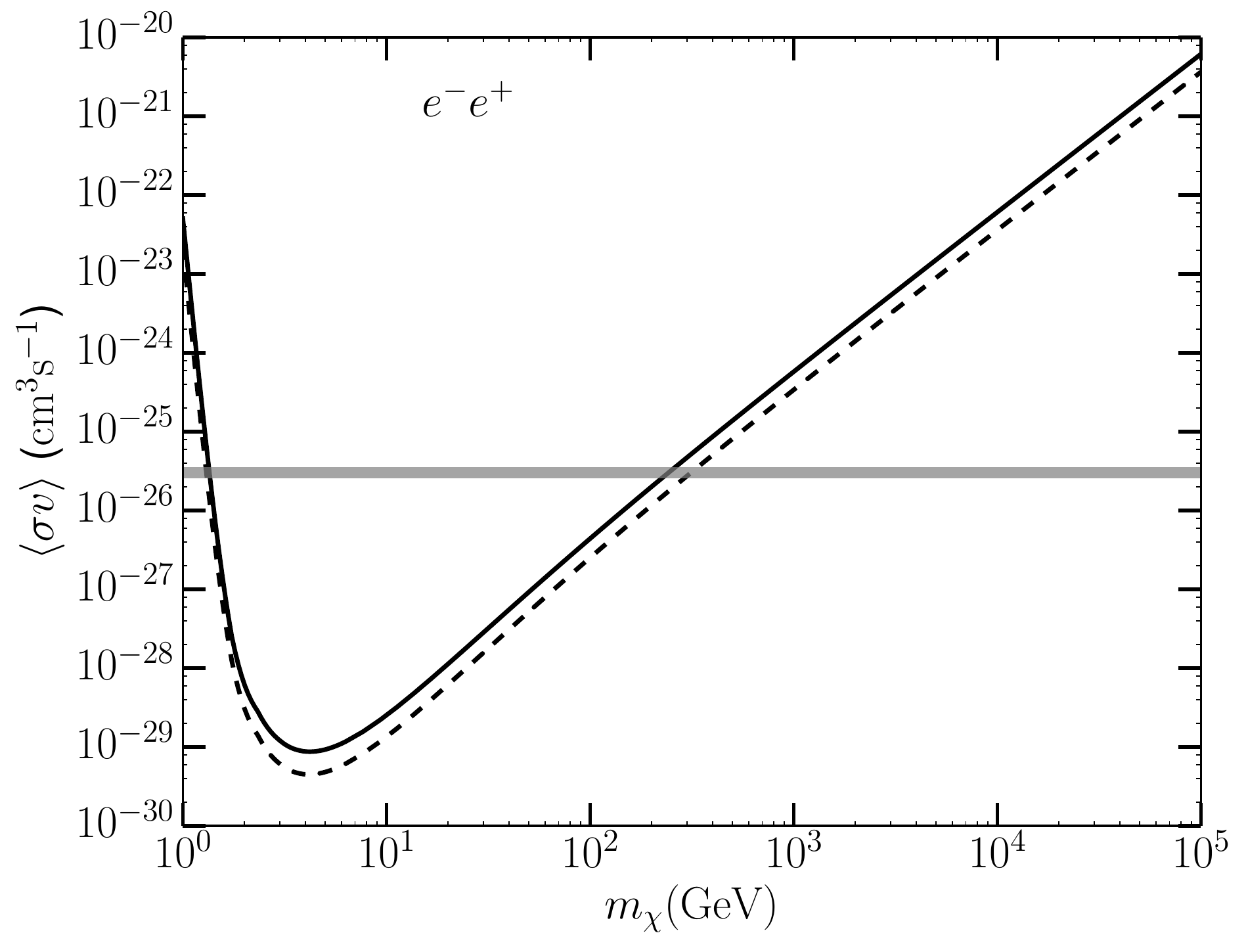}
    \caption{Limits on the annihilation cross section that result from considering an Einasto profile for the SC (dashed). Comparison with the corresponding limits from Fig. \ref{pic:limits} (hard line).} \label{pic:Ein}
  \end{center}
\end{figure}

Fig. \ref{pic:Ein} shows the comparison between our limits on the annihilation cross section into $b\bar b$ and $e^-e^+$ with the ones that result from considering an Einasto distribution \eqref{eq:Einasto}. 
The same parameters for the magnetic field and diffusion length as in Fig.~\ref{pic:limits} were used here.

As expected, considering different density functions yields curves that behave in the same manner, only normalized differently. The ratio of both curves is (approximately) equal to the ratio of their corresponding J-factors (see e.~g. eq. \eqref{eq:flux}).

\section{Diffusion coefficient uncertainties}
\label{sec:app2}
\begin{figure}[!ht]
  \begin{center}
\includegraphics[width=.49\linewidth,natwidth=610,natheight=642]{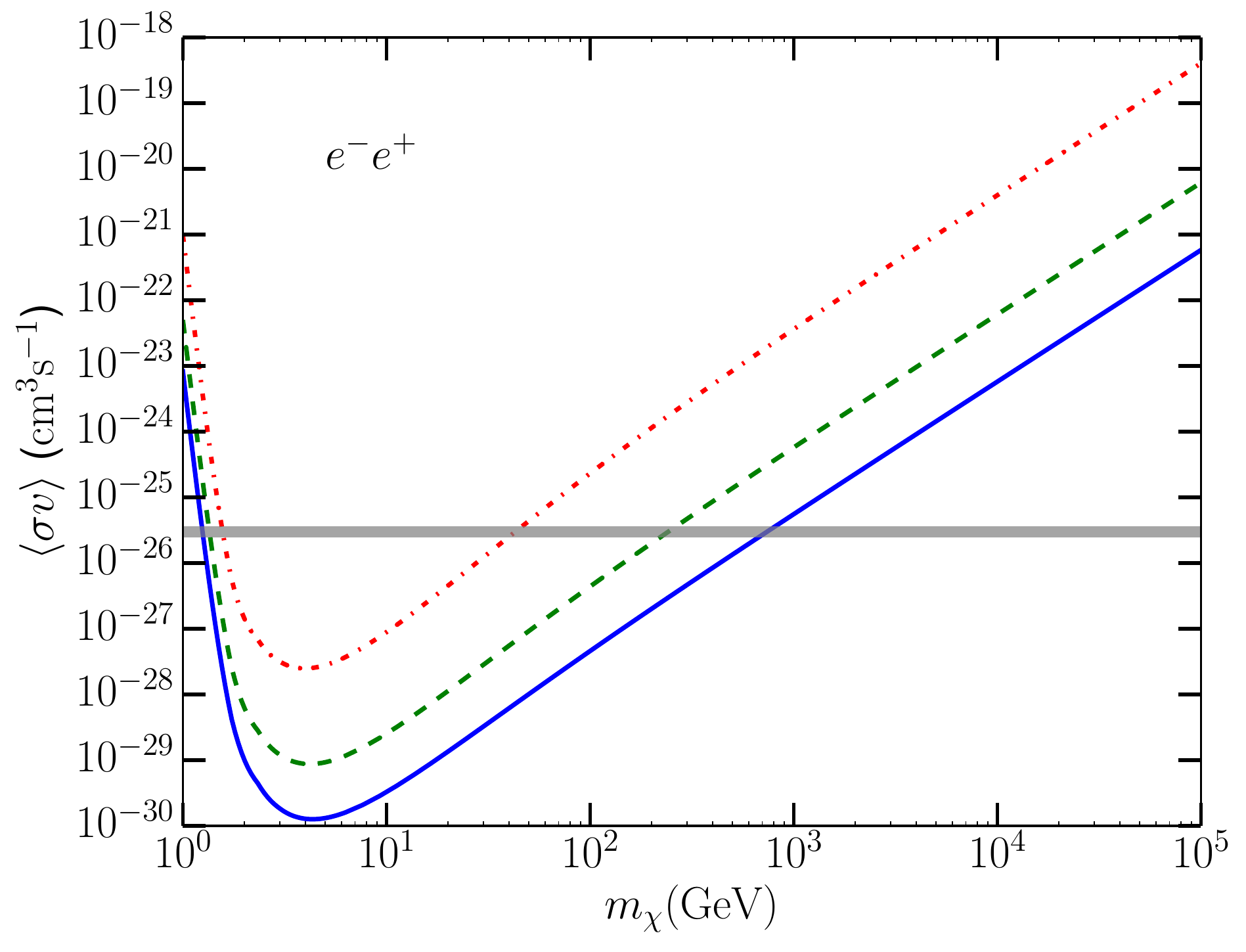} 
\includegraphics[width=.49\linewidth,natwidth=610,natheight=642]{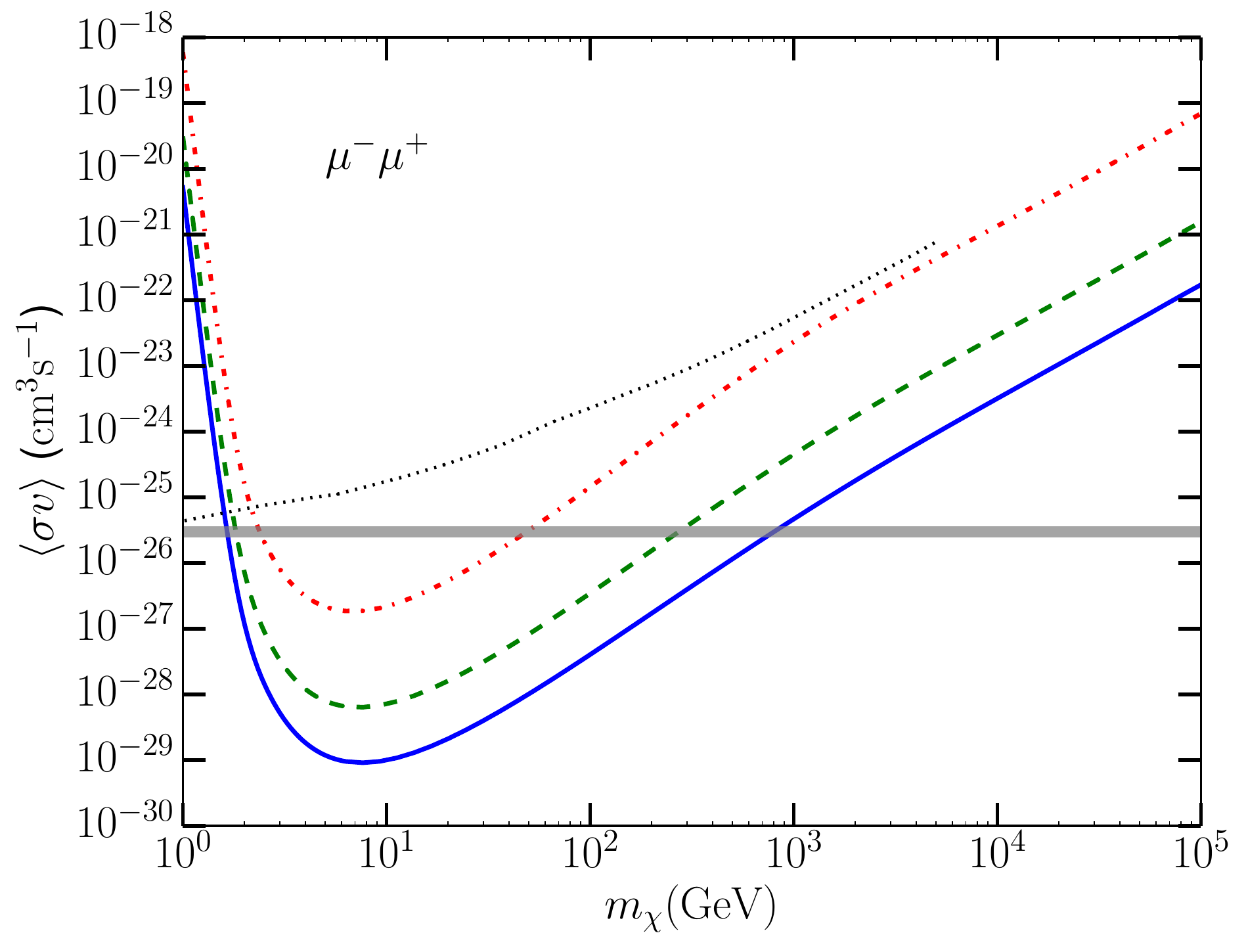} \includegraphics[width=.49\linewidth,natwidth=610,natheight=642]{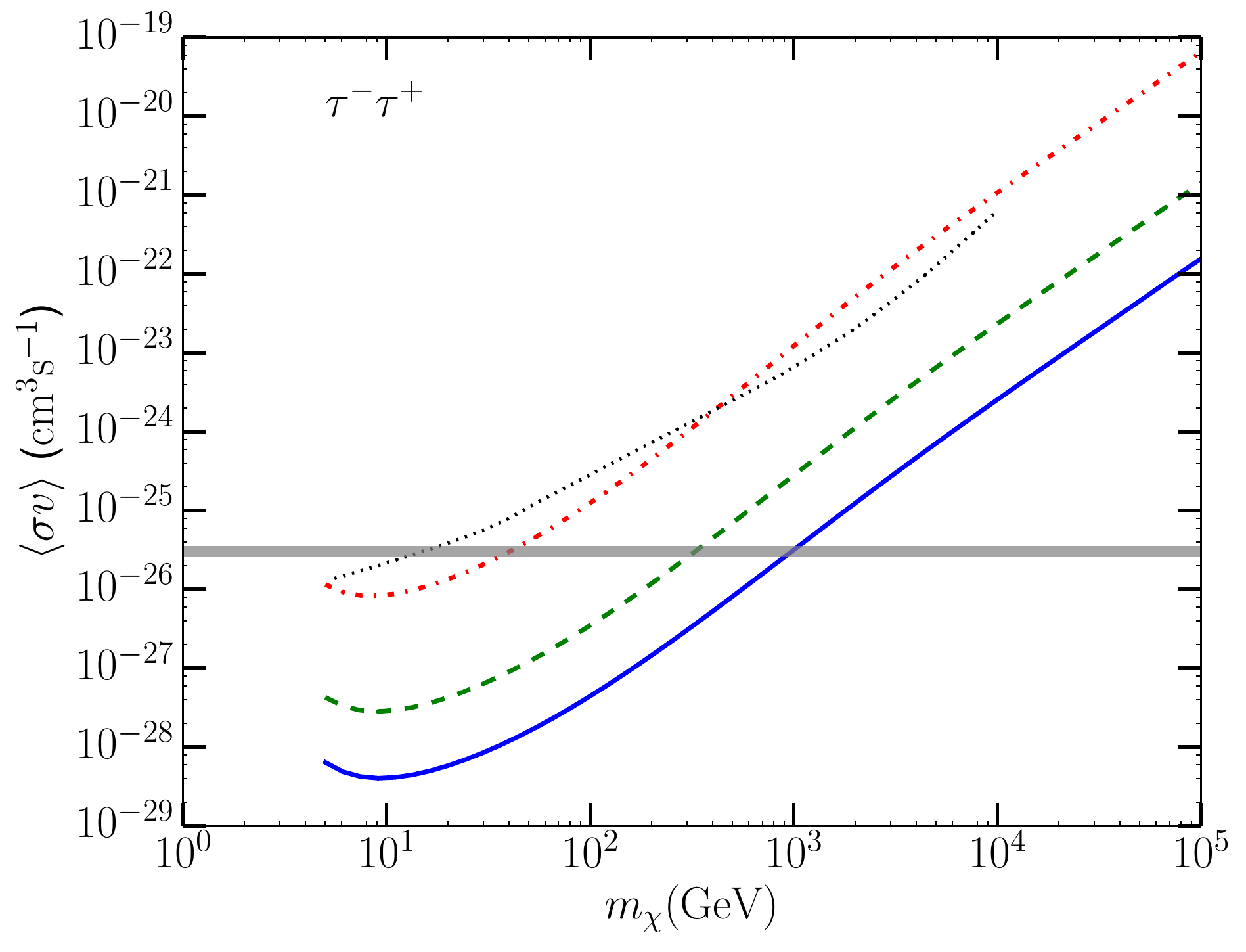} \includegraphics[width=.49\linewidth,natwidth=610,natheight=642]{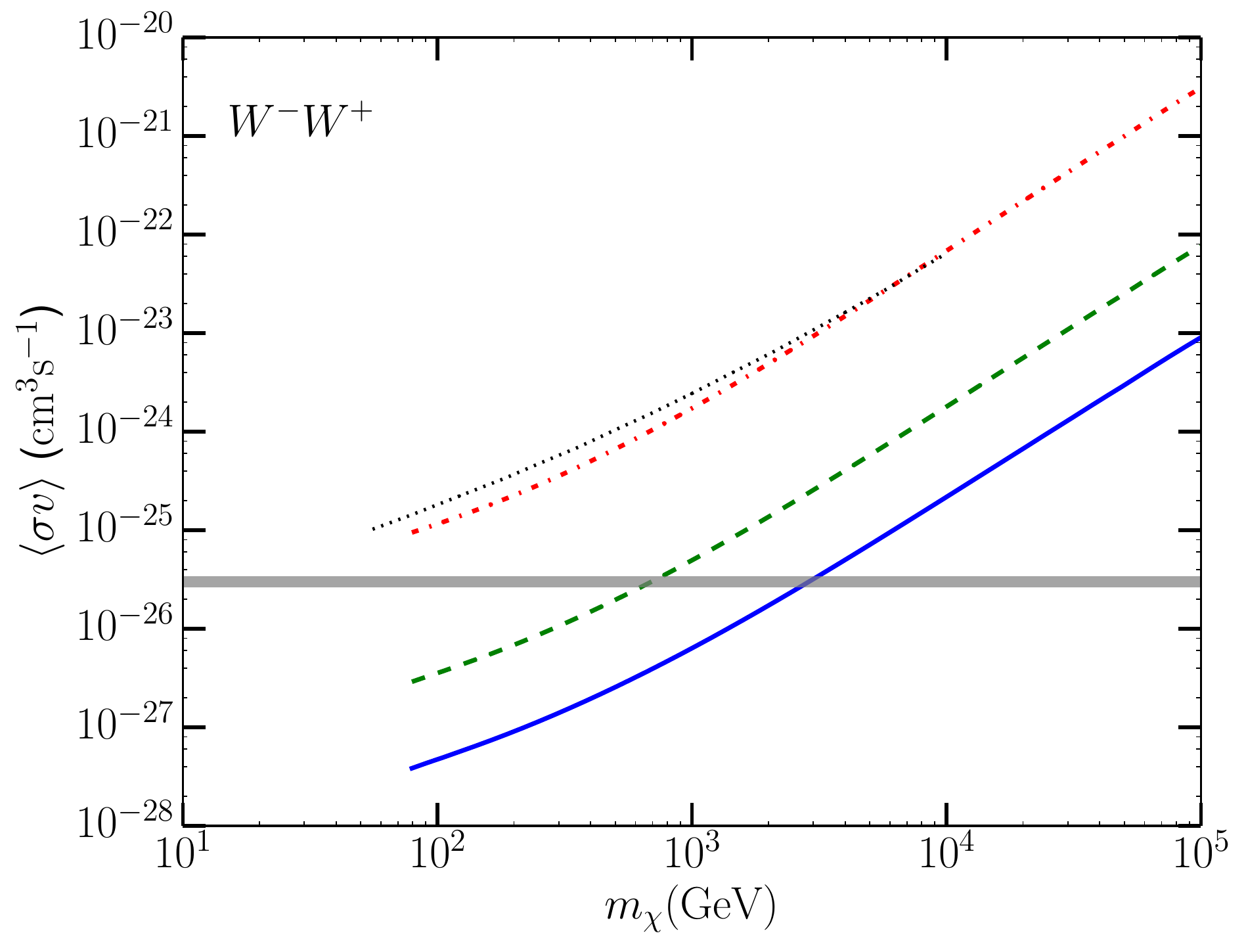}  \includegraphics[width=.49\linewidth,natwidth=610,natheight=642]{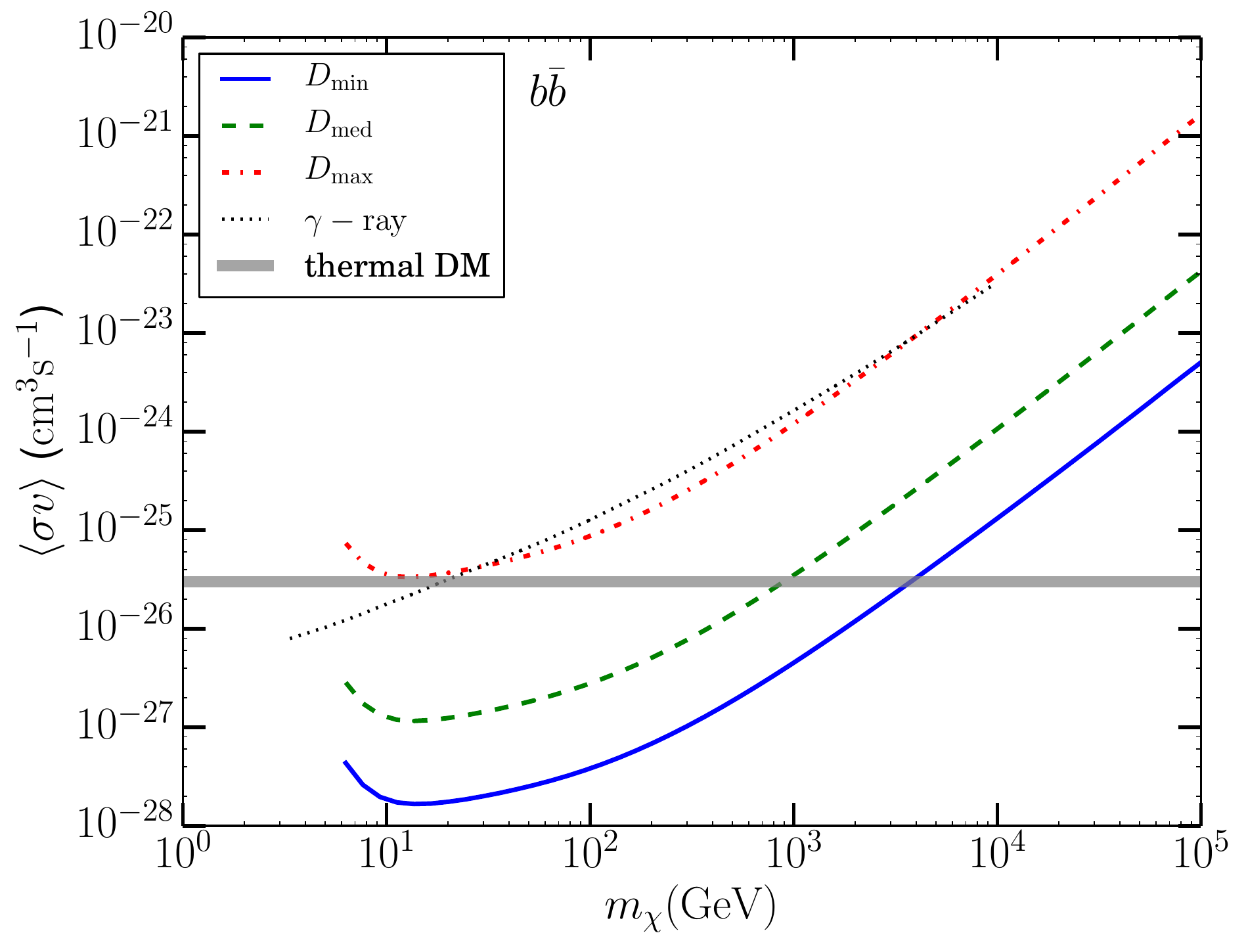}
    \caption{\label{fig:minmedmax} Limits on the annihilation cross section for 'maximum', 'medium' and 'mininum' values of the normalization $D_0$ in a Kolmogorov diffusion model. Comparison with the limits drawn by  Ref. \cite{Drlica-Wagner:2014yca} using gamma rays.} \label{pic:Diffcoef}
  \end{center}
\end{figure}

The limits shown in Fig. \ref{pic:limits} were obtained assuming the normalization $D_0=10^{27}$ cm$^{2}$s$^{-1}$ for the diffusion coefficient, which, as explained in the text, corresponds to a typical loss-diffusion length of $\sim$500~pc. 
To account for the uncertainty of its value, we show in Fig.~\ref{pic:Diffcoef} the impact of varying the diffusion coefficient over the range $D_0=4\times10^{25}-10^{29}$ ~cm$^{2}$s$^{-1}$, corresponding to $\sqrt{\lambda_0}=100~\text{pc}-5~\text{kpc}$.

The range considered for $D_0$ is larger than the typically accepted for Galactic cosmic ray propagation. 
In particular, data analyses that consider the ratio of primary (e.~g. boron, antiprotons) to secondary (e.~g. carbon, protons) cosmic ray fluxes prefer normalizations of the diffusion coefficient within $D_0=4.83\times10^{26}-2.31\times10^{28}$~cm$^{2}$s$^{-1}$ \cite{Donato:2003xg}.

The limits shown in fig. \ref{fig:minmedmax} become weaker as $D_0$ is increased. 
This is precisely the opposite behaviour to the situation where cosmic-ray data is used to put constraints on the annihilation cross section of DM.
There, the 'MIN' constraints are the weakest while the limits that result from the 'MAX' model are the strongest.

This is expected then the farther the electrons/positrons diffusively propagate, the higher the probability that they would ``hit'' Earth (corresponding to larger cosmic-ray fluxes). 
On the other hand the synchrotron emission becomes less intense as the diffusion grows because the emission will be less and less localized.
In the opposite case, the electrons are 'trapped' in a smaller volume and the emission is more intense.

When comparing with the constraints set by gamma ray studies, as done in Fig. \ref{pic:limits}, we conclude that our synchrotron predictions yield interesting comparable annihilation cross sections. 

\bibliographystyle{JHEP}
\providecommand{\href}[2]{#2}
\begingroup\raggedright

\endgroup
\end{document}